\documentclass[]{revtex4-1}

\usepackage{amsmath}
\usepackage{amssymb}
\usepackage[caption=false]{subfig}
\usepackage[dvipsnames]{xcolor}
\PassOptionsToPackage{caption=false}{subfig}
\usepackage{epsf}
\usepackage[final]{graphicx}
\usepackage{sidecap}
\sidecaptionvpos{figure}{c}
\usepackage{amsmath,amssymb,bm,color,mathrsfs,wasysym}
\usepackage{epsfig}
\usepackage{amsfonts}
\usepackage{cancel}
\usepackage{ulem}
\usepackage{float}
\usepackage{tikz}
\usepackage[left=2cm,right=4cm,marginparwidth=3cm,]{geometry}

\newcounter{saveeqn}

\newcommand{\be}{\begin{equation}}
	\newcommand{\ee}{\end{equation}}
\newcommand{\bea}{\begin{eqnarray}}
	\newcommand{\eea}{\end{eqnarray}}
\newcommand{\bi}{\begin{itemize}}
	\newcommand{\ei}{\end{itemize}}

\definecolor{LightBlue}{rgb}{0.0, 0.0, 1.0}

    \makeatletter
\def\@fnsymbol#1{\ensuremath{\ifcase#1\or \dagger\or *\or
   \mathsection\or \mathparagraph\or \|\or **\or \dagger\dagger
   \or \ddagger\ddagger \else\@ctrerr\fi}}
    \makeatother

\begin{document}



\title{Shear dynamics of confined membranes}

\thanks{Electronic Supplementary Information (ESI) available: simulation movies. See DOI: 10.1039/D1SM00322D}

\author{Thomas Le Goff}
\affiliation{Aix-Marseille Univ, CNRS, IBDM, Turing Centre for Living System, Marseille, France.}

\author{Tung B.T. To}
\affiliation{Instituto de F\'{i}sica, Universidade Federal Fluminense, Avenida Litor\^{a}nea s/n, 24210-340 Niter\'{o}i RJ, Brazil.}

\author{Olivier Pierre-Louis}
\affiliation{Institut Lumi\`ere Mati\`ere, UMR5306 Universit\'e Lyon 1-CNRS, Universit\'e de Lyon, 69622 Villeurbanne, France.}
\email{Olivier.Pierre-Louis@univ-lyon1.fr}

\date{\today}

\begin{abstract}
We model the nonlinear response of a 
lubricated contact composed of a two-dimensional lipid membrane
immersed in a simple fluid between two parallel flat and porous walls
under shear. 
The nonlinear dynamics of the membrane gives rise to a 
rich dynamical behavior depending on the shear velocity.
In quiescent conditions (i.e., absence of shear), the membrane 
freezes into a disordered labyrinthine
wrinkle pattern. We determine the wavelength of this pattern
as a function of the excess area of the membrane
for a fairly general form of the confinement potential
using a sine-profile ansatz for the wrinkles.
In the presence of shear, we find four different
regimes depending on the shear rate.
Regime I. 
For small shear, the labyrinthine pattern is still frozen,
but exhibits a small drift which is mainly along the shear direction. 
In this regime, the tangential forces on the walls due to the presence of the 
membrane increase linearly with the shear rate.
Regime II.
When the shear rate is increased above a critical value, the membrane rearranges,
and wrinkles start to align along the shear direction. This regime
is accompanied by a sharp drop of the tangential forces on the wall.
The membrane usually reaches a steady-state configuration
drifting with a small constant velocity
at long times. However, we also rarely observe oscillatory dynamics in this regime.
Regime III. 
For larger shear rates, the wrinkles align strongly along the shear direction,
with a set of dislocation defects which assemble in pairs. The tangential forces
are then controlled by the number of dislocations, and by the number
of wrinkles between the two dislocations within each dislocation pairs.
In this dislocation-dominated regime, the tangential forces in the 
transverse direction most often
exceed those in the shear direction.
Regime IV. 
For even larger shear, the membrane organizes into 
a perfect array of parallel stripes with no defects. 
The wavelength of the wrinkles is still identical to the wavelength
in the absence of shear. In this final regime, 
the tangential forces due to the membrane vanish.
These behaviors give rise to a non-linear rheological behavior
of lubricated contacts containing membranes.
\end{abstract}

\maketitle

\section{Introduction}
\label{sec:intro}

Biolubrication is vital for the function of biological organs of living species.
One of the key ingredients of biolubricating systems is lipid membranes,
which reduce the friction coefficient 
and wear of rubbing surfaces \cite{JeanPaulRieu2008}. 
An example of biolubrication system is synovial fluid confined between opposing cartilage surfaces 
within the joints of human and other animals. 
Synovial fluid has three important components: surface-active phospholipids, hyaluronan (acid) and proteoglycan-4 (proteins) \cite{Schmidt2007}. 
In synovial fluids, surface-active phospholipids form stacks of parallel bilayers at the surface of the cartilage, 
which reduce the friction coefficient \cite{Loison2015}.  Swann \textit{et al.} \cite{Swann1984} 
suggest that the response of synovial fluids to friction is related to joint diseases such as degeneration, traumatic injury, inflammation, infection, acute gout syndrome, and arthritis. In another study for knee joint, Tadmor \textit{et al.} \cite{Tadmor2002} proposed that the combination of synovial membrane, ligament, tendon and skin acts together like a spring between opposing cartilages and applies tensile force on the joint as one lifts his or her leg. In contrast the presence of hyaluronan helps to slow down the compression between cartilages as one puts weight on the joint.

The role of parallel lipid bilayers to reduce the friction coefficient at biological surfaces such as cartilage has prompted several groups to cover model artificial surfaces with lipid bilayers, thereby reducing the friction coefficient to very low values around $10^{-3}$ \cite{JeanPaulRieu2008}. 
These results suggest novel routes towards applications in the automotive and manufacturing industry where more effective and efficient lubrications are in high demand in order to save fuel, increase engine durability, and reduce environmental pollution. These industrial criteria of lubrication systems may be achieved by the development and consumption of low friction materials, coatings, and lubricants \cite{Erdemir2005}. In addition, the study of friction in model lubrication system may have implications for the lubrication of biomedical devices and microelectromechanical systems and suggests applications in living systems \cite{Raviv2003,Briscoe2006}.

In the past decades, many studies have focused on the dynamics of stacks of membranes under shear (see, e.g., Ref.\cite{Gov2004} and references therein).
They have pointed out the instabilities of these stacks induced by shear,
and the stabilizing effect or shear on thermal fluctuations of the membranes\cite{Olmstead2002}.
In this paper we investigate a simple situation where a single membrane sandwiched between two porous walls.
We focus on wrinkling patterns formed by membranes due to their
excess area, and show that the nonlinear dynamics of these wrinkles gives rise
to a rich rheological behavior.
Wrinkle patterns have been observed in many systems where thin films are present such as stressed thin films on soft substrate \cite{Huang2006}, 
confined liquid membranes \cite{membraneadhesion2D2018},
stretched polyethylene sheets \cite{Cerda2003} or 
confined biogel membranes \cite{LeocmachSciAdv2015}. 
In the first two cases \cite{Huang2006,membraneadhesion2D2018}, the thin film forms sinusoidal wavelike patterns 
with a characteristic wavelength and the wrinkles meander to form the isotropic labyrinthine patterns \cite{LeBerre2002}. 

Our work extends our previous study in one dimension~\cite{PierreLouis2017}. 
Here, we consider a lubricated contact containing
a two-dimensional membrane immersed in a Newtonian fluid between 
two flat permeable walls.
These permeable walls account for porous
materials such as the collagen of the cartilage matrix in joints,
or the cytoskeleton which is in contact with membranes in biological cells.
The walls move with constant and opposite velocities leading to a shear flow.
We find that, while confinement of the membrane gives rise
to the formation of wrinkles that store the excess
area, shear can rearrange these wrinkles. 
The membrane then experiences nontrivial
configurations, which lead to a back-flow that produces
tangential forces on the confining walls.
These tangential forces exhibit a nonlinear dependence
on the shear rate, and therefore give rise to a complex 
nonlinear rheological response of the lubricated contact.

In the following, we first present the lubrication model in Section 2.
We consider a two-dimensional inextensible membrane with bending rigidity
in a simple liquid between two flat and permeable walls.
We describe the wall-substrate with a generic membrane-wall potential that diverges as the membrane approaches the wall,
thereby preventing the membrane from crossing the walls
(this is in contrast with harmonic potentials, such as that used e.g. in Ref.~\cite{Gov2004}).

In section 3, we investigate the dynamics in the absence of shear.
We find that the wrinkles relax to a  steady-state
composed of a labyrinthine pattern of wrinkles.
A sine-ansatz profile provides a quantitative
prediction for the width of the wrinkles.

Section 4 is devoted to the numerical 
investigation of the membrane dynamics in the presence of shear, and of the
resulting tangential forces acting on the walls. We find
four different regimes as a function of the
shear rate. In regime I, the membrane exhibits a 
labyrinthine pattern that is similar to that found
in the absence of shear. However, the membrane exhibits
a small drift, mainly oriented along the shear
direction. The forces due to the presence of the 
membrane then increase linearly with the shear rate.
When the shear rate exceeds a critical value $\upsilon_c$,
the membranes start to rearrange, and align partially
along the shear direction. We denote this regime as Regime II.
In regime II, the forces on each wall drop sharply. 
We have also sometimes observed oscillatory dynamics
in regime II.
Further increase of the shear rate leads to 
regime III, where wrinkles are
strongly ordered along the shear direction.
In regime III, dislocations can be observed.
These dislocations form pairs. 
In this regime, the tangential forces exhibit a complex
dependence on the shear rate, and 
the tangential forces in the transverse direction
are found to be often larger than in the shear direction.
Finally, for very large shear rate, the 
dislocations disappear and the membrane
exhibits a perfect array of wrinkles parallel to
the shear direction. In this regime,
hereafter denoted as regime IV, the tangential forces vanish.

In section 5, we determine the critical velocity
$\upsilon_c$ from a balance between the shear term and 
the other terms in the dynamical equations.
Section 6 is devoted to the analysis of the tangential
force on the walls. 
For small shear rate in regime I, we use the sine anstaz to
obtain a quantitative prediction of the tangential forces.
Furthermore, we show that the total force
acting on the walls can be decomposed
as a sum of contributions of dislocation pairs in regime III.
In Section 7, we claim that the transition to
regime IV is controlled by finite size effects.
Finally, a summary and discussion of our results
are presented in Section 8.

\section{Model}
\subsection{Evolution equation for a confined membrane under shear}

\begin{figure}
\begin{center}
    {\includegraphics[width=0.5\textwidth]{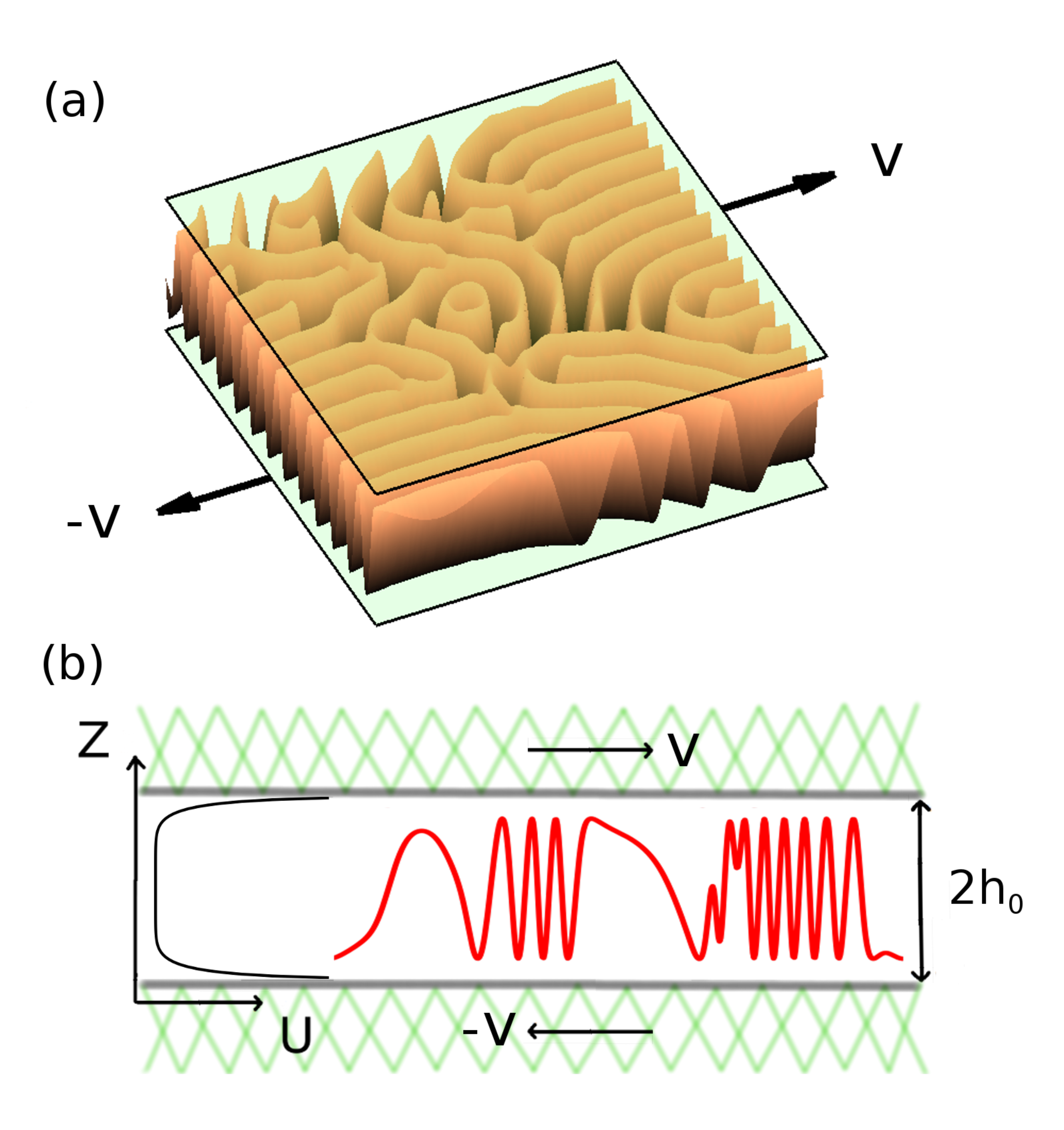}}
\caption{
Membrane between two repulsive substrates. For the sake of clarity, the vertical scale is strongly magnified in (a) and (b) (our analysis applies in the limit where the vertical scale is much smaller than the horizontal scale, i.e., when slopes are small). Profiles in (a) and (b) are obtained from simulation with $\Delta A_*=0.1503$, $V=2$ at $T=4\times 10^5$. (a) 3D view; (b) 2D view, left black curve: membrane-walls interaction potential, right red curve: a section of the membrane profile.
}
\label{fig:schematic}
\end{center}
\end{figure}

In previous papers, we have contributed to the modeling
of the dynamics of membranes placed between two walls, with~\cite{PierreLouis2017}
or without~\cite{membraneadhesion2D2018} shear.
Beyond the presence of shear induced by the motion of the walls,
two main physical ingredients of the models can be used to
categorize the dynamics that we have investigated so far.

The first ingredient is the interaction of the 
membrane with the walls. The interaction of the membrane with one wall
can be dominated by purely repulsive interactions,
such as hydration interactions~\cite{Swain2001}, or repulsion 
due to polymer brushes~\cite{Sengupta2010}.
When attractive forces are present, such as those
induced at long range by van der Waals force~\cite{Israelachvili2015}, 
or at short distances by molecular binders~~\cite{Sengupta2010}, one obtains a potential well for the membrane
at a distance $h_w$ from a wall.
If the walls are attractive and if the distance between the two walls
is larger than $h_w$, then the potential experienced by the membrane exhibits
a double-well profile. In contrast, if the interaction is purely
repulsive or if the interaction is attractive but the distance
between the two walls is smaller than or similar to $2h_w$, then the membrane
experiences a single-well potential. Single and double-well potentials
give rise to different membrane behaviors. For example in the absence of shear,
coarsening of adhesion domains coexisting with wrinkles can be observed with double-well potentials
as discussed in Ref.~\cite{membraneadhesion2D2018}, while 
only frozen labyrinthine wrinkle patterns can be seen with single-well potentials,
as discussed below.

The second ingredient is the balance between two dissipation mechanisms,
controlled by the viscosity of the liquid and the permeability of the walls.
The relative roles of the liquid flow through the walls and along the walls
is characterized by a dimensionless number~\cite{membraneadhesion2D2018}
\begin{equation}
\label{eq:barnu_def}
    \bar{\nu}=12\frac{\kappa^{1/2}\mu\nu}{\mathcal{U}_0^{1/2}h_0^2}.
\end{equation}
Here, $\kappa$ is the bending rigidity of the membrane, 
$\mu$ is the viscosity of the fluid surrounding the membrane, 
$\nu$ is a kinetic coefficient describing the permeability of the wall, 
$\mathcal{U}_0$ is an energy scale 
for the adhesion potential, and $2h_0$ is the distance
between the two walls. 
In this work we focus on the limit of very large permeability  $\bar{\nu}\gg 1$,
but where viscosity and hydrodynamic flows along the membrane are still relevant. 
This limit which aims at describing membranes sandwiched between porous biological substrates such as collagen or the cytoskeleton,
is discussed quantitatively in more details in section \ref{sec:sumresul} and Ref.~\cite{membraneadhesion2D2018}.
The opposite limit of impermeable walls $\bar{\nu}\ll 1$ is suitable to describe,
e.g., substrates covered by other lipid membranes as discussed 
in Ref.~\cite{membraneadhesion2D2018}. Models which account for finite
permeability $\bar{\nu}\sim 1$ have also been reported in Ref.~\cite{membraneadhesion2D2018}.

The effect of shear on membrane dynamics 
with impermeable walls and a double-well potential has been investigated 
in our previous work in Ref.~\cite{PierreLouis2017} within a one-dimensional model. In this work,
we focus on the case of permeable walls with a single-well potential within
a two-dimensional model.

More precisely, we consider the dynamics of a two-dimensional membrane
confined between two walls at height $\pm h_0$ moving at constant
velocities $\pm \upsilon_0$.
A three-dimensional representation of the membrane and the walls is shown in Fig. \ref{fig:schematic}(a and b).
The walls are permeable with a 
permeability constant $\nu$. Within the small-slope approximation, the evolution equation for the membrane was derived in 
one dimension with shear in Ref.\cite{PierreLouis2017}, and in two dimensions
without shear in  Ref.\cite{membraneadhesion2D2018}. The combination of these
models to obtain a two-dimensional model with shear is straightforward.
Here, we do not report the full derivation, but we motivate the 
different terms appearing in the equations.

In the limit of large permeability, i.e. when $\bar\nu$ is large, 
an equation governing the membrane height $h(x,y)$ can be obtained from the lubrication
(small slope) limit. The main ingredients of the derivation are reported in 
Appendix~\ref{secapp:lubric_force}, while more details are provided in Ref.~\cite{membraneadhesion2D2018}.
This equation takes a simple form 
\begin{equation}
\label{eq:dynamic2D}
\partial_t h=\frac{\nu}{2}f_z-\frac{\upsilon_0}{h_0}h\partial_xh,
\end{equation}
where $x$-axis is parallel to the shear direction and $f_z$ accounts for the internal forces on the membrane
along the $z$ direction orthogonal to the walls
\begin{equation}
f_z=-\kappa \Delta^2h+\sigma_0 \Delta h-{\cal U}^\prime(h),
\end{equation}
where $\kappa$ is the bending rigidity~\cite{Canham1970, Helfrich1973} of the membrane. 
The term $\sigma_0$ is a nonlocal tension which enforces a constant
membrane area.  Membrane tension usually accounts for the entropic tension of the membrane due to thermal
fluctuations\cite{Lipowsky1995,Seifert1995}, as well as some possible amount of finite membrane extensibility.
Following Refs.\cite{Young2014,membraneadhesion2D2018,PierreLouis2017}, the tension here
emerges to leading order in the dynamics of the membrane at small slopes
as a result of the constraint of local inextensibility. 
Inserting the  membrane excess area in the small slope limit
\begin{align}
\label{eq:excess_area_def}
\Delta {\cal A}&= \int_0^{{\cal L}_x} dx \int_0^{{\cal L}_y} dy 
\frac{1}{2} (\nabla h)^2,
\end{align}
into Eq. (\ref{eq:dynamic2D}), 
the membrane area conservation condition $\partial_t \Delta {\cal A}=0$ provides the expression of 
the  tension of the membrane
\begin{equation}
\sigma_0 = 
\frac{\int_0^{{\cal L}_x} dx \int_0^{{\cal L}_y} dy 
\left\{ \left[ \frac{\nu}{2} \left( \kappa \Delta^2 h + {\cal U}'(h) \right) 
+ (\upsilon_0/h_0) h\partial_xh \right] \Delta h \right\}}
{\int_0^{{\cal L}_x} dx \int_0^{{\cal L}_y} dy \left\{ \frac{\nu}{2}\left( \Delta h \right) ^2 \right\}}.
\end{equation}

Finally, ${\cal U}$ is the confinement potential, i.e., 
the free energy for placing the membrane at the height $h$.
This potential accounts for the interaction of the membrane with the porous walls.
The walls account for
the confinement of the membrane induced by the cytoskeleton of the cell \cite{Sheetz2001,Speck2012}, 
or a biological substrate \cite{Maciver1992, Berrier2007}, 
or other membranes~\cite{Braga2002, Asfaw2006}. 
We use a generic potential of the form
\begin{align}
\label{eq:potential_def}
{\cal U}(h)= \frac{{\cal U}_0}{[1-(h/h_0)^p]^m}
\end{align}
with $m,p>0$, and $p$ even. 
Such a confinement potential allows one to account simultaneously for 
a divergence at the walls ${\cal U}(h)\approx {\cal U}_0p^{-m}(1\pm(h/h_0))^{-m}$
when $h\rightarrow \pm h_0$
with an arbitrary power $m$,
and for a small amplitude behaviour 
${\cal U}(h)\approx {\cal U}_0+m{\cal U}_0(h/h_0)^p$ when $h\rightarrow 0$
with arbitrary power $p$.

\subsection{Forces of the walls}

The expression of the tangential forces on the walls generalizes 
the results of Ref.~\cite{PierreLouis2017} derived within a one-dimensional model
in the limit of small slopes. 
These forces originate in the shear stress exerted by the fluid on the walls.
For two-dimensional membranes, the force on each wall is opposite to the force 
on the other wall. 
A detailed derivation of these forces is reported is Appendix \ref{secapp:lubric_force}.
The two components of the force per unit area on one wall along the direction $x$ and $y$ are
\begin{align}
\label{eq:totalforce2D}
&f_{wx}=\frac{\mu}{h_0} \upsilon_0 + f_{\mathrm{mem},x}
&f_{wy}=f_{\mathrm{mem},y}
\end{align} 
The first contribution $\upsilon_0 \mu/h_0$ is the viscous friction force
due to the simple shear of the fluid
along $x$ in the absence of membrane. The second contribution is due to the presence of the membrane
\begin{align}
\label{eq:membrane_force_2D}
&f_{\mathrm{mem},x}=\frac{1}{2{\cal L}_x {\cal L}_y} \int_0^{{\cal L}_x} dx \int_0^{{\cal L}_y} dy \left\{ \frac{h_0}{2} \left( 1-\frac{h^2}{h_0^2} \right) \partial_x f_z \right\}, \nonumber \\
&f_{\mathrm{mem},y}=\frac{1}{2{\cal L}_x {\cal L}_y} \int_0^{{\cal L}_x} dx \int_0^{{\cal L}_y} dy \left\{ \frac{h_0}{2} \left( 1-\frac{h^2}{h_0^2} \right) \partial_y f_z \right\}. 
\end{align}
A qualitative discussion of these expressions follows.
In the lubrication limit, each flow above or below
the membrane is a Poiseuille flow parallel to the walls $\upsilon_{x,y}$. 
These flows contribute to the tangential friction forces between the two walls when they
produce viscous shear stresses $-\mu\partial_z\upsilon_{x,y}|_{h= h_0}$ on the upper wall
and  $+\mu\partial_z\upsilon_{x,y}|_{h=- h_0}$ on the lower wall which exhibit opposite signs,
i.e., when $\partial_z\upsilon_{x,y}|_{h= h_0}$ and $\partial_z\upsilon_{x,y}|_{h=- h_0}$
have the same sign. This corresponds typically to an antisymmetric flow, 
with the property $\upsilon_{x,y}(z)=-\upsilon_{x,y}(-z)$. 
The first antisymmetric contribution to the flow is 
the  average shear flow $\upsilon_{x,y}(z)=\upsilon_0 z/h_0$
due to the imposed motion of the walls. This contribution gives rise to the first term in Eq. (\ref{eq:totalforce2D}).
In addition, the membrane normal force $f_z$ produces a difference
of pressure between the fluids above and below the membrane.
Spatial variations of $f_z$ therefore produce pressure gradients that give rise to additional fluid flow.
This additional fluid flow is at the origin of the contributions $f_{\mathrm{mem},x}$
and $f_{\mathrm{mem},y}$ in Eq. (\ref{eq:membrane_force_2D}).
At this point, two important remarks should be made.
First, the average flow on one side of the membrane 
vanishes when the membrane approaches the wall, due to an increase of viscous dissipation.
Second, each flow above and below the membrane is a Poiseuille flow which cannot be antisymmetric
by itself.
As a consequence of these two properties, (i)
the most antisymmetric flow is produced by placing the membrane in the middle of the cell
and (ii) when the membrane is placed close to one wall, the flow is then essentially a symmetric Poiseuille.
This is at the origin of the factor $1-h^2/h_0^2$ in Eq. (\ref{eq:membrane_force_2D}),
which is maximum when $h=0$ and vanishes when $h\rightarrow\pm h_0$.

Following the same lines as in Eq. (19) of Ref. \cite{PierreLouis2017}, 
we use periodic boundary conditions to rewrite the membrane contribution to the forces as
\begin{align}
\label{eq:memforce2D}
&f_{\mathrm{mem},x}=\frac{1}{2h_0{\cal L}_x {\cal L}_y} \int_0^{{\cal L}_x} dx \int_0^{{\cal L}_y} dy (h\partial_xhf_z), \nonumber \\
&f_{\mathrm{mem},y}=\frac{1}{2h_0{\cal L}_x {\cal L}_y} \int_0^{{\cal L}_x} dx \int_0^{{\cal L}_y} dy (h\partial_yhf_z). 
\end{align}

\subsection{Normalization}\label{sub:normlaization}

We define the normalized  time $T=t/t_0$,
where
\begin{equation}\label{eq:normalisationtime}
t_0=\frac{2h_0^2}{\nu\mathcal{U}_0},
\end{equation}
the normalized height $H=h/h_0$,
and the normalized space variables parallel to the walls $(X,Y)=(x,y)/\ell_\parallel$,
where
\begin{equation}\label{eq:normalisationdistance}
\ell_\parallel=\left(\frac{\kappa h_0^2}{\mathcal{U}_0}\right)^{1/4}
\end{equation}
accounts for the lengthscale in the $(x,y)$ plane on which a membrane 
placed away from the center between the walls decays back to the center
in the linear approximation, i.e. in a harmonic potential.
Since we consider the lubrication limit (small slopes), we assume that 
\begin{equation}\label{eq:lubricationlimit}
\epsilon=\frac{h_0}{\ell_\parallel}\ll1.
\end{equation}

\begin{figure}
\begin{center}
    {\includegraphics[width=0.4\textwidth]{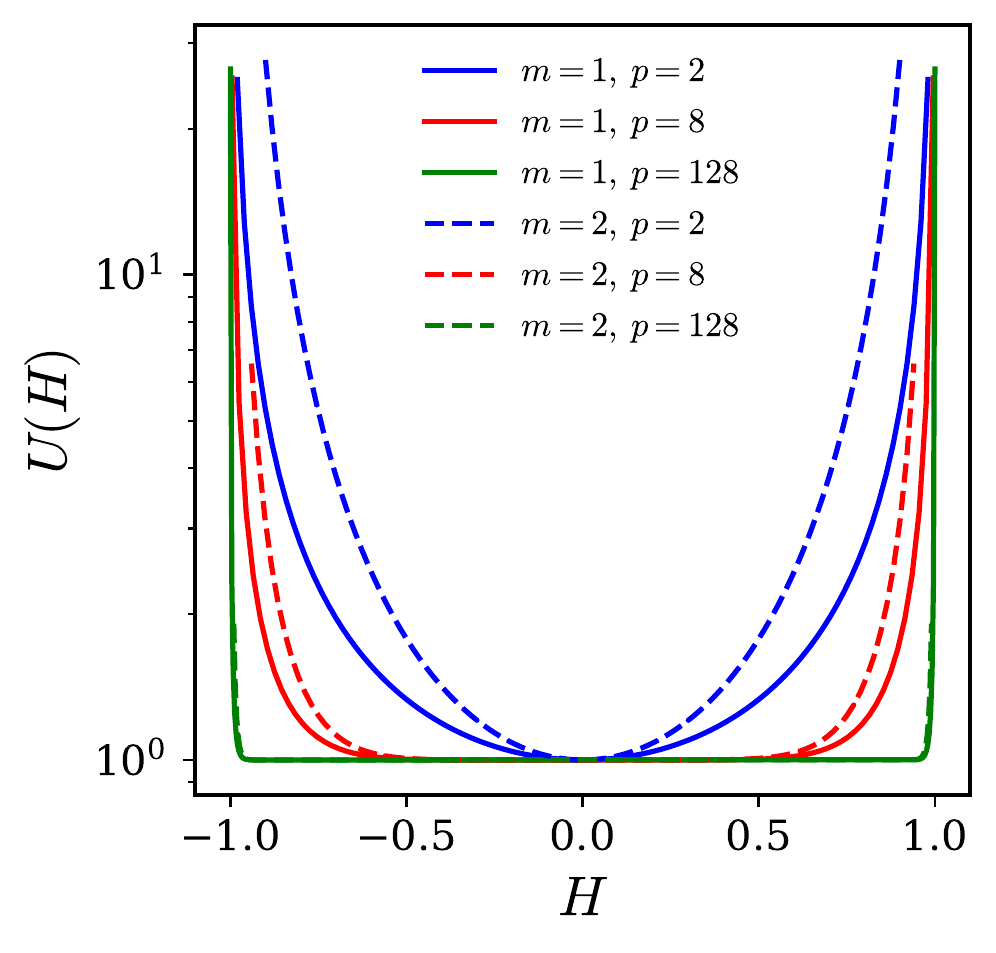}}
\caption{
Confinement potential Eq. (\ref{eq:potential_def_normalized}).
At small $H$, the potential can be varied from a harmonic profile
for $p=2$ to a square potential for larger $p$.
Close to the walls, the divergence of the potential is controlled by $m$.
}
\label{fig:potentials}
\end{center}
\end{figure}

We also define the normalized shear velocity
\begin{equation}
V=\frac{\upsilon_0}{\upsilon_{\parallel}},
\end{equation}
where
\begin{equation}\label{eq:normalisationvelocity}
\upsilon_{\parallel}=\frac{\ell_\parallel}{t_0}=\frac{\nu\mathcal{U}_0^{3/4}\kappa^{1/4}}{2h_0^{3/2}}.
\end{equation}
This leads to the normalized evolution equation
\begin{equation}
\label{eq:dynamic2Dnormalized}
\partial_TH=F_Z-VH\partial_XH.
\end{equation}
The normalized forces on the membrane then read
\begin{equation}
F_Z=-\Delta^2H+\Sigma_0\Delta H-U'(H),
\end{equation}
where
\begin{equation}
\label{eq:sigma0_normalized}
\Sigma_0=\frac{\int_0^{L_X} dX \int_0^{L_Y} dY \left\{ \left[\left( \Delta^2 H + U'(H) \right) + V H\partial_XH \right] \Delta H \right\}}{\int_0^{L_X} dX \int_0^{L_Y} dY \left\{\left( \Delta H \right) ^2 \right\}},
\end{equation}
and the normalized potential reads
\begin{align}
\label{eq:potential_def_normalized}
U(H)=\frac{1}{(1-H^p)^m}.
\end{align}
The changes in the shape of the normalized potential are reported in Fig. \ref{fig:potentials}.

While analytical results will be discussed for arbitrary 
values of $p$ and $m$, we have focused on the 
case $m=1$ and $p=8$ in the numerical simulations.
This rather large value of $p$ is chosen
in order to mimic a square-like potential. This will
allow us to check our predictions far beyond the harmonic regime.

The force on the membrane along the $Z$ direction is normalized as $F_Z=(h_0/\mathcal{U}_0)f_z$. 
In normalized coordinates, $L_X$ (resp. $L_Y$) represents the length of the system in the $X$ (resp. $Y$) direction in normalized variables. 

For a given confinement potential, the dynamics is controlled by two dimensionless parameters:
the normalized density of excess area
\begin{equation}
\Delta A_*=\frac{1}{\epsilon^2} \, \frac{\Delta\mathcal{A}}{\mathcal{L}_x\mathcal{L}_y}=\frac{\Delta A}{L_XL_Y}
\end{equation}
and  the normalized shear velocity $V$.

We also define the normalized forces on each wall as 
\begin{align}
F_{wX}&=\frac{24\kappa^{1/4}h_0^{1/2}}{\mathcal{U}_0^{5/4}}f_{wx}
=\bar{\nu} V+12\langle H\partial_XHF_Z \rangle,\nonumber \\
F_{wY}&=\frac{24\kappa^{1/4}h_0^{1/2}}{\mathcal{U}_0^{5/4}}f_{wy}
=12\langle H\partial_YHF_Z \rangle,
\label{eq:normalizedtotalforce}
\end{align}
and we have introduced the spatial averaging notation for any function $\psi$
\begin{equation}
\label{eq:avrgdef}
\langle \psi \rangle=\frac{1}{L_XL_Y} \int_0^{L_X}dX\int_0^{L_Y}dY \psi.
\end{equation}
Thus, the two contributions of the membrane to the tangential forces on each wall read
\begin{equation}
\label{eq:memforce2Dnormalized}
F_{\mathrm{mem},X}=12 \langle H\partial_XHF_Z \rangle, \quad
F_{\mathrm{mem},Y}=12 \langle H\partial_YHF_Z \rangle.
\end{equation}


\section{Dynamics of confined membranes without shear}
\subsection{Frozen labyrinthine states}

\begin{figure*}
\begin{center}
    {\includegraphics[width=\textwidth]{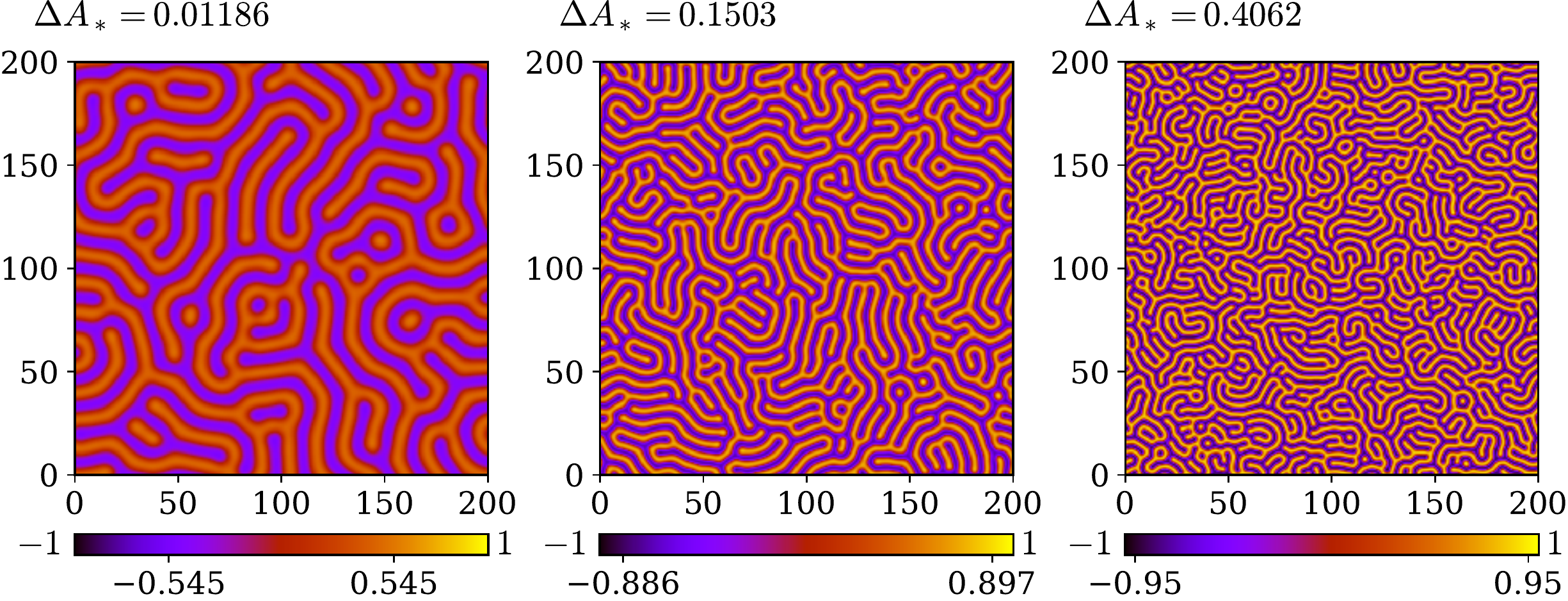}}
\caption{Labyrinthine patterns in quiescent conditions.
Top-views of the membrane height in steady state for several membrane excess areas in quiescent conditions ($V=0$). 
The marks on the color bars show the minimum and maximum membrane heights in each image.
}
\label{fig:noshear_topview}
\end{center}
\end{figure*}

In the absence of shear $V=0$ and starting from random initial conditions 
the membrane dynamics relaxes quickly to a frozen isotropic labyrinthine pattern. 
In this regime, the tension is always negative, as expected for a membrane
in compression due to confinement with a fixed excess area.
We use a numerical scheme described in Ref.\cite{membraneadhesion2D2018} which preserves precisely
the membrane area, and random initial conditions as described in Appendix~\ref{a:random_init_cond}.
Fig. \ref{fig:noshear_topview} shows topviews of the membrane height profile 
for several membrane excess area. 
The minimum and maximum membrane height are shown as marks on the color bars. 
A visual inspection of the topviews shows that the wrinkle wavelength $\lambda$ 
decreases and the maximum membrane height $\lvert h\rvert_\mathrm{max}$ increases as 
the excess area $\Delta A_*$ is increased.
Simulation movies for $V=0$ are shown in Electronic Supplementary Information (ESI)\dag ({\it Movie 1}).

\subsection{Sine-ansatz}

In the absence of shear, the last term of Eq. (\ref{eq:dynamic2D}) vanishes, and
our model shares similarities with previous models
described in the literature, 
which leads to the formation of wrinkles.
For example, the model for wrinkle formation in stressed films on
soft substrates reported in Ref.\cite{Huang2006}, initially
presents some amount of coarsening, where the wavelength
of the wrinkles increases with time, and ultimately leads to
a frozen labyrinthine pattern of wrinkles in the presence of equiaxial stress.
Labyrinthine patterns have also been reported 
in the literature for the Swift-Hohenberg equation by Le Berre \textit{et al.} \cite{LeBerre2002}.
However, our equations differ from these models and exhibit two main specific features.
First, our non-local tension is constant in space and depends on time via Eq. (\ref{eq:excess_area_def}), while it is a 
local tensorial stress in Ref.\cite{Huang2006}, and a time-independent constant in the Swift-Hohenberg equation.
Second, our confinement potential is different and diverges at the walls.

We have also recently reported on the observation of labyrinthine patterns 
or endless coarsening in
the dynamics of lipid membranes subject to a double-well adhesion potential 
~\cite{membraneadhesion2D2018}.
Following the same lines as in our previous study~\cite{membraneadhesion2D2018}, 
we relate the wrinkle wavelength $\lambda$ and $\Delta A_*$ analytically using an ansatz that assumes a sinusoidal membrane profile 
\begin{equation}
\label{eq:sineansatz_def}
H(\zeta)=a \cos(q\zeta),
\end{equation}
where $a$ and $q$ are two positive constants, 
and $\zeta$ is the space variable locally orthogonal to 
the wrinkle orientation in the $X,Y$ plane.
Such a sinusoidal membrane profile can be seen in Fig. \ref{fig:schematic}(b). 
The red curve shows a cross section of the membrane profile by a vertical plane.
We see that a part of the curve on the right is almost periodic and sinusoidal.
In this part, the cross section direction is almost perpendicular 
to the direction of the membrane wrinkles.
In the other parts of the curve, where the cross section is not 
perpendicular to the direction of the wrinkles, the membrane profile looks 
more random, sometimes with flatter parts when the section is aligned
with the top of the crests, or the bottom of the valleys of the wrinkles.
Substitution of Eq. (\ref{eq:sineansatz_def}) into Eq. (\ref{eq:excess_area_def}) 
leads to a simple relation between $q$ and $\Delta A_*$ 
\begin{equation}
\label{eq:sineansatz_noshear}
\Delta A_*=\frac{1}{4}q^2 a^2=\frac{\pi^2}{\lambda^2} a^2,
\end{equation}
where $\lambda=2\pi/q$ is the wrinkle wavelength.
This geometrical relation was already reported
in previous work\cite{Cerda2003,membraneadhesion2D2018}.
Within the sine ansatz the root-mean-square corrugation of the membrane height $\langle H^2 \rangle$ 
is proportional to the amplitude $\langle H^2 \rangle^\frac{1}{2}=a/\sqrt{2}$. 

\begin{figure}
\begin{center}
    {\includegraphics[width=0.9\textwidth]{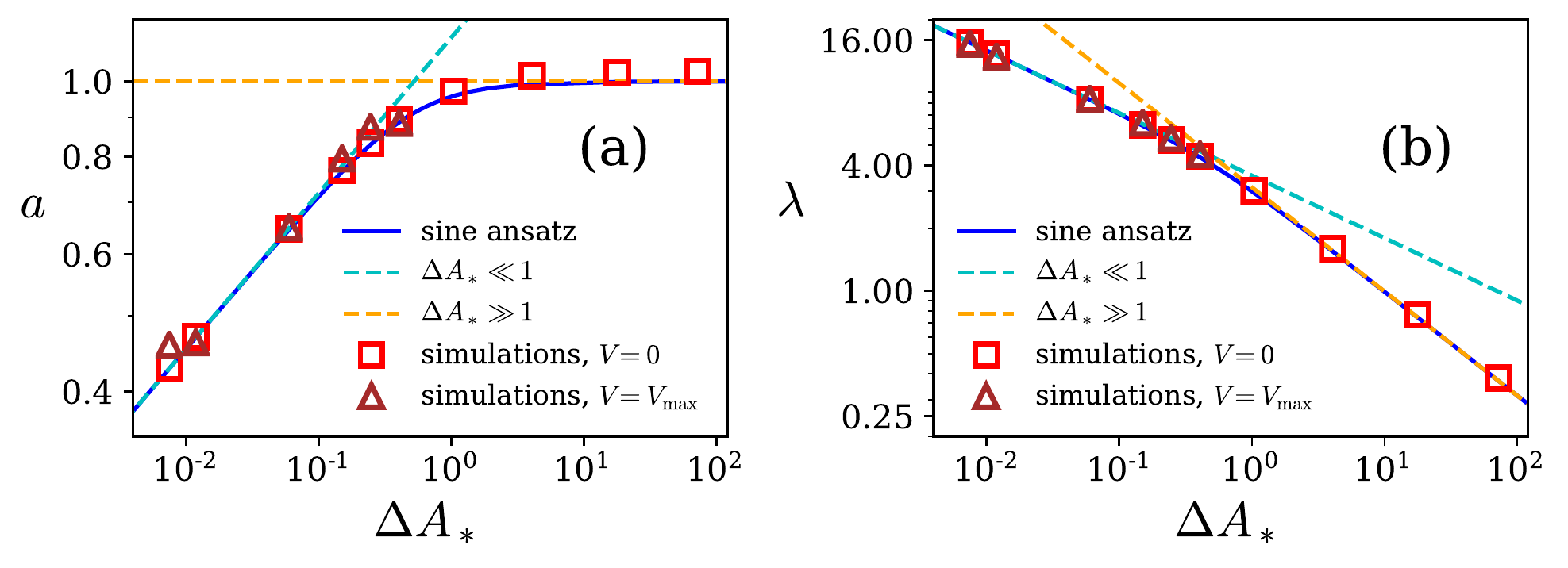}}
\caption{Sine-ansatz predictions of wrinkle size.
(a) Wrinkle amplitude $a$ and (b) wrinkle wavelength $\lambda$
are plotted as a functions of the membrane excess area $\Delta A_*$.
Blue solid curves: full analytical results from the sine ansatz (see Appendix \ref{secapp:sineansatz_noshear}). 
Cyan dashed lines: analytical results in limit of small $\Delta A_*$ in Eq. (\ref{eq:general_analytical_noshear}),
leading to $\lambda \simeq 3.5765 \Delta A_*^{-\frac{3}{10}}$, 
$a \simeq 1.1385 \Delta A_*^\frac{1}{5}$ for $m=1$ and $p=8$.
Yellow dashed lines: analytical results in limit of large $\Delta A_*$: $a=1$ and $\lambda=\pi \Delta A_*^{-1/2}$. 
Red squares: simulation results for $V=0$. Brown triangles: simulation results for periodic state 
under shear at $V = V_\mathrm{max}$.
} 
\label{fig:scalings_noshear}
\end{center}
\end{figure}

A second relation between $q$ and $a$
is obtained from the minimization of the total energy,
including bending rigidity and interaction potential,
for a fixed excess area. 
Below, we report on the limit of small and large excess area.
The details of the general calculation are reported in
Appendix \ref{secapp:sineansatz_noshear}.

In the limit of small excess areas $\Delta A_* \ll 1$, the result takes a simple form
\begin{equation}
\label{eq:general_analytical_noshear}
\lambda =  \pi C_{m,p} \, \Delta A_* ^{-\frac{1}{2}+\frac{2}{p+2}}, 
\quad\quad a= C_{m,p}\,  \Delta A_*^\frac{2}{p+2},
\end{equation}
where the expression of the constant $C_{m,p}$ is given in Eq. (\ref{eq:Cmp}).
In this small amplitude regime, two limiting
cases are in order. First, when $p=2$ 
(shown as blue curves in Fig. \ref{fig:potentials}), 
the potential is harmonic for small $h$
and the wavelength $\lambda$ is independent of $\Delta A_*$. This means that
the wrinkle lengthscale is identical to the scaling lengthscale $\ell_\parallel$.
In addition, we have $a\sim \Delta A_*^{1/2}$.
Second, the opposite case of large $p\gg 1$ corresponds to a square-like potential which is
very flat between the two walls 
(shown as green curves in Fig. \ref{fig:potentials}). 
In this square-like potential, $\lambda\rightarrow \pi \Delta A_*^{-1/2}$
and the amplitude $a\rightarrow 1$ is independent of $\Delta A_*$. Hence, the excess area
is stored by an increase of amplitude at fixed wavelength when $p=2$,
and by a decrease of wavelength at fixed amplitude when $p\gg 1$.

For large excess area $\Delta A_* \gg 1$, the membrane amplitude approaches 
the walls so that $a\approx 1$. In this case, the sine ansatz in Eq. (\ref{eq:sineansatz_noshear}) 
implies that the wavelength $\lambda$ is independent of $m$ and $p$,
\begin{equation}
\label{eq:lambda_Alarge}
\lambda=\pi\Delta A_*^{-\frac{1}{2}}, 
\quad\quad a=1.
\end{equation}
Interestingly, we notice the small excess area expansion for large $p\gg 1$
catches quantitatively the limit of large excess area.

The predictions of the sine-ansatz and the related small and large excess area limits 
for the potential Eq. (\ref{eq:potential_def_normalized}) with $m=1,p=8$ 
are in quantitative agreement with the
full simulations, as shown in Fig. \ref{fig:scalings_noshear}.
Note that the sine anstaz discards the meandering and branching
of wrinkles in labyrinthine patterns.
The accuracy of the prediction of the sine ansatz therefore indicates 
that meandering and branching play a negligible
role in the section of the wavelength.

\section{Dynamics of confined membrane under shear: simulation results}

In this Section we present the simulation results 
in the presence of shear, when $V\neq0$. 
All simulations are started with random initial conditions.

\subsection{Steady-drifting states and Oscillatory states}
\label{sec:steady_oscillatory}

\begin{figure}
\begin{center}
{\includegraphics[width=0.7\textwidth]{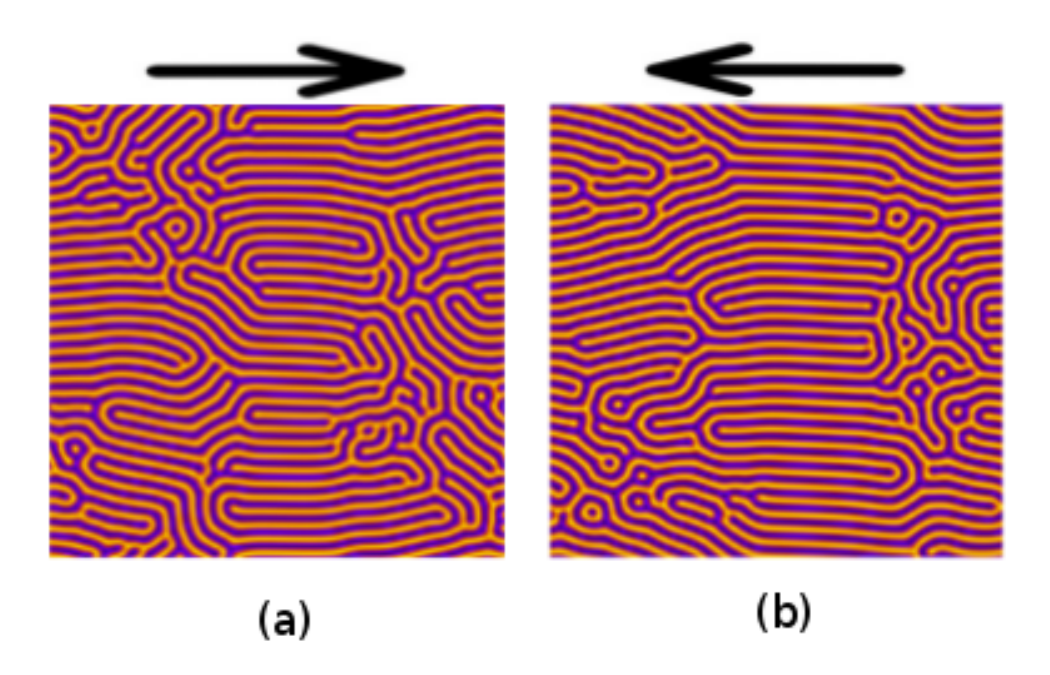}}
\caption{
Drift direction.
Membrane profiles for $\Delta A_*=0.1503$ and $V=2$ and different initial conditions with opposite drifting directions along $X$ to the right (a) and to the left (b).
}
\label{fig:driftdirection}
\end{center}
\end{figure}

In most simulations, and independently from the value
of the shear rate $V$, the membrane reaches a constant profile at long times
which drifts with a constant velocity $\mathbf V_d$ with non-vanishing components
$V_{dX}$ and $V_{dY}$ along $X$ and $Y$.
Some steady-state configurations together with the direction
of their drifts in the $X$ direction are shown in Fig. \ref{fig:driftdirection}.
Simulation movies for $V_{dX}>0$, $V_{dX}<0$ and $V_{dX}\approx 0$ are shown in ESI\dag ({\it Movie 2, Movie 3 {\rm and} Movie 4}, respectively).
The method for the measurements
of the drift velocities, and a table summarizing their values are reported in Appendix \ref{secapp:Vdriftapp}.
In addition, as in the quiescent case, the tension is always negative.

Since the dynamical equation (\ref{eq:dynamic2D})
is invariant under the variable changes $Y\rightarrow -Y$, and 
$(X,H)\rightarrow (-X,-H)$, there is no preferred drift direction.
The drift can therefore be seen as a consequence of the
random asymmetry of the disordered steady-states.
More precisely, drifting steady-state profiles obey
\begin{align}
\label{eq:driftsteady_main}
    -V_{dX}\partial_XH-V_{dY}\partial_YH=-\Delta^2H+\Sigma_0\Delta H -U'(H)-VH\partial_XH.
\end{align}
 Hence, for each steady-state $H(X,Y)$ with drift velocity $(V_{dX},V_{dY})$,
 there is a second steady-state  $H(X,-Y)$ with drift velocity $(V_{dX},-V_{dY})$,
 and a third steady-state $-H(-X,Y)$ with drift velocity  $(-V_{dX},V_{dY})$.
As a consequence, although a non-zero drift can be present in a given simulation,
the drift averaged over many simulations started with random initial conditions 
 should vanish.
 Indeed we observe in the simulations that the 
signs of $V_{dX}$ and $V_{dY}$ can be positive and negative with 
roughly equal probabilities.

\begin{figure*}
\begin{center}
{\includegraphics[width=\textwidth]{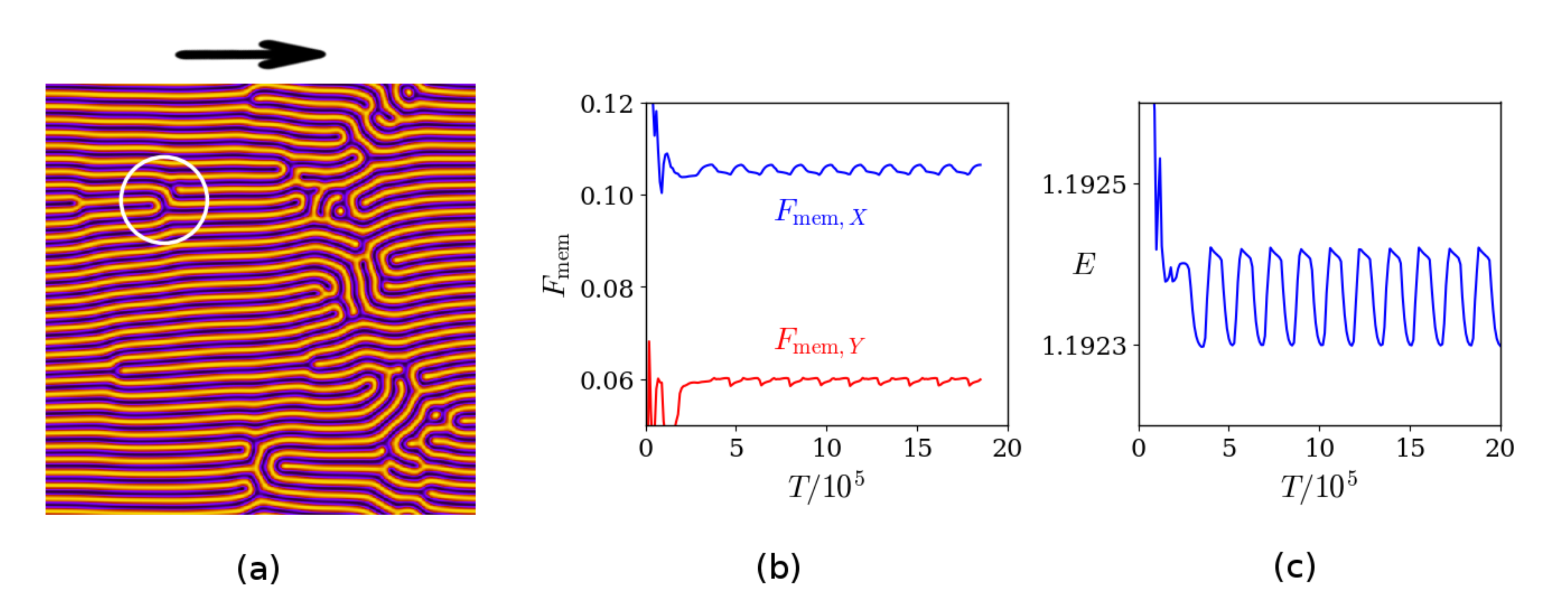}}
\caption{
Oscillations. (a) Membrane pattern drifting to the right with an oscillating pair of defects (indicated by the white circle) for $\Delta A_*=0.1503$ and $V=3$.
(b) Corresponding components of the membrane force acting on each wall $F_\mathrm{mem}$
. 
(c) Evolution of the corresponding membrane energy $E$ from Eq. (\ref{eq:energynormalized}).
}
\label{fig:oscillating_ripples}
\end{center}
\end{figure*}

Note that, although the final state is a steady-drifting state
in the vast majority of cases, this is not always the case. 
Indeed, for some values of $\Delta A_*$ and $V$ we found patterns that oscillate periodically. 
As an example we show  
the membrane profile for $\Delta A_*=0.1503$ and $V=3$
in Fig. \ref{fig:oscillating_ripples}(a). 
In this simulation, there is a
pair of defects (shown by the white circle) that moves alternately forwards and backwards
along the shear direction $x$. We define a defect as the end of the line following the top of a wrinkle crest
or the bottom of a wrinkle valley
(more example of defects will be provided below in Fig. \ref{fig:patternA0p1503}).
This oscillation is superimposed to a global drift.
Simulation movie related to this case is shown in ESI\dag ({\it Movie 5}).

The existence of drifting or oscillatory states can be related to the non-variational character
of the shear term $VH\partial_XH$ in  Eq. (\ref{eq:dynamic2Dnormalized}).
Indeed, as discussed in Ref.\cite{membraneadhesion2D2018}, 
in the absence of this term, the dynamics is decreasing
the free energy of the membrane, comprising bending energy and potential energy
\begin{equation}
\label{eq:energynormalized}
    E=\left\langle \frac{1}{2} (\Delta H)^2+U(H) \right\rangle.
\end{equation}
Hence, $\partial_T E< 0$ when $\partial_TH\neq 0$.
Such monotonic decrease of the energy implies that the system cannot
evolve and go back to the same state, which would
correspond to the same energy: drift and oscillations are forbidden.
This strong constraint is lost in the presence of non vanishing shear.

\begin{figure}
\begin{center}
    {\includegraphics[width=0.45\textwidth]{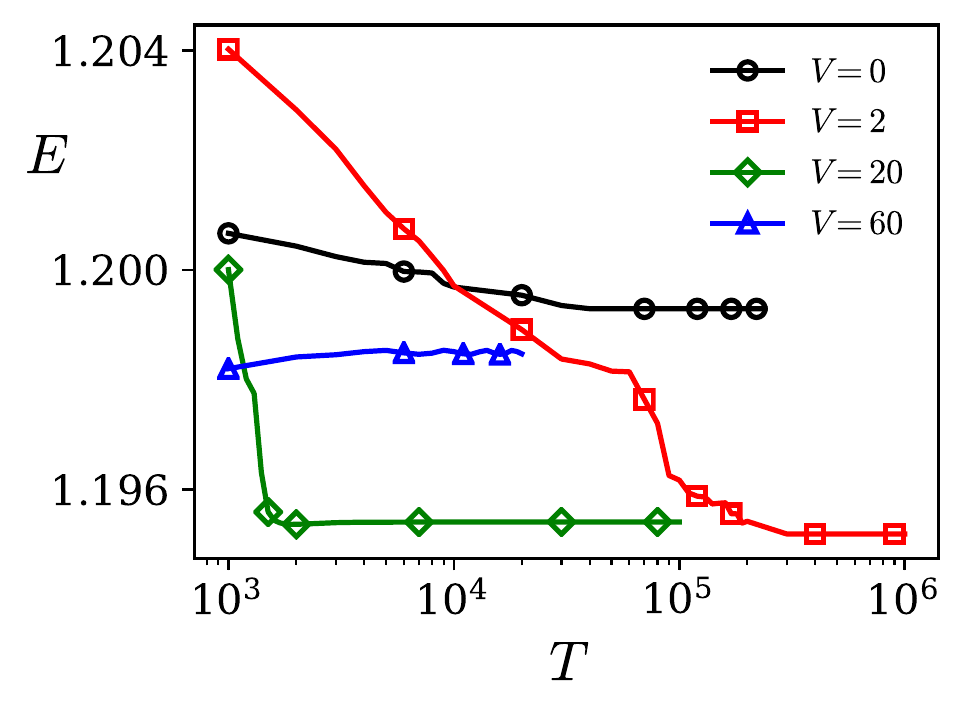}}
\caption{
Evolution of the membrane energy $E$ with time $T$ when the final state is
a steady-state with a constant drift. All simulations are performed at $\Delta A_*=0.1503$.
}
\label{fig:Esteady}
\end{center}
\end{figure}

The time evolution of the energy $E$ for some drifting steady-states
in Fig. \ref{fig:Esteady} shows that 
in some cases the membrane energy can increase with time when $V\neq0$. 
In addition, while the energy $E$ ultimately reaches a constant value in drifting steady-state 
as shown in Fig. \ref{fig:Esteady}, 
it oscillates in the periodic oscillating-drifting state as shown in Fig. \ref{fig:oscillating_ripples}(c).

\subsection{Dynamical regimes as a function of the shear rate $V$}

\begin{figure*}
\begin{center}
{\includegraphics[width=\textwidth]{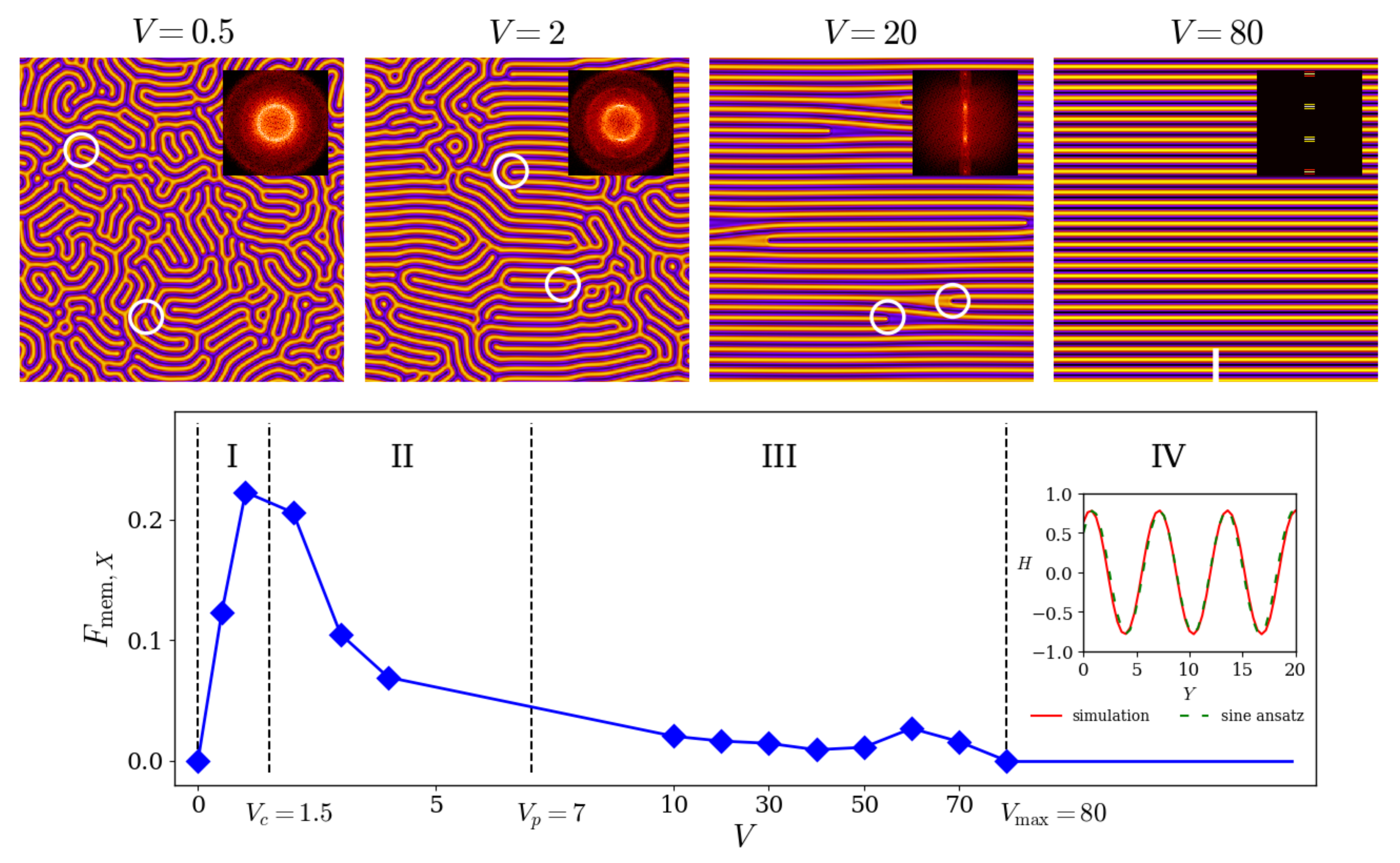}}
\caption{
Steady-state regimes of confined membranes under shear, and related tangential forces on the walls.
All simulation profiles and forces are obtained for $\Delta A_*=0.1503$.
Top panels: The membrane profile exhibits four different regimes as the shear velocity is varied $V$ in simulations.
White circles highlight some representative defects.
Insets (top panels): zooms of the Fourier transforms in log scale with $\lvert q_x \rvert < 2\pi/5$ and $\lvert q_y \rvert < 2\pi/5$; for $V=80$, $\lvert q_x \rvert <\pi/50$.
Bottom panel: variation of the membrane force along $X$ on each wall $F_{\mathrm{mem},X}$ as the function of the shear rate $V$.
In order to show Regimes I and II, the interval from $V=0$ to $V=10$ is magnified in the abscissa. 
Inset (bottom panel):  cross-section of the membrane profile in Regime IV from simulation
along the white vertical line of the top view at $V=80$.
The profile is very close to the prediction of the sinusoidal ansatz  Eq.~(\ref{eq:sineansatz_def},\ref{eq:general_analytical_noshear}). 
The same regimes, with similar membrane profiles and similar variations of the force, 
are observed for different values of $\Delta A_*$, as shown in Fig. \ref{fig:totalforce}. 
}
\label{fig:patternA0p1503}
\end{center}
\end{figure*}

The dependence of the membrane patterns on the shear velocity $V$ 
exhibits 4 different regimes. Typical configuration
corresponding to these regimes are presented in the top panels in Fig. \ref{fig:patternA0p1503}.

{\bf Regime I.} At small shear velocities, the membrane presents an isotropic labyrinthine pattern 
which is very similar to the pattern found in the absence of shear.
This isotropy is clearly seen in the Fourier transform on the first panel in Fig. \ref{fig:patternA0p1503}.
This pattern also exhibits the same wavelength as in the absence of shear,
as shown in Fig. \ref{fig:scalings_noshear}(b).
The pattern exhibits a random drift which increases
in amplitude when $V$ increases. The dift is also anisotropic,
with a larger amplitude along the $X$ axis,
i.e. $|V_{dX}|>|V_{dY}|$. 

{\bf Regime II.} When the shear rate $V$ exceeds a critical value $V_c$, the
membrane starts to reorganize and becomes anisotropic, with 
wrinkles partially aligned in the $X$ direction.
The Fourier transform in the second figure on the top right of Fig. \ref{fig:patternA0p1503} presents a
clear anisotropy.
This is the only regime where  oscillatory states have been observed.

{\bf Regime III.} For larger shear rates $V>V_p$, the membranes form a parallel
array of wrinkles with dislocations. There are 4 types of dislocations,
which can be deduced from each other via the $X\rightarrow -X$,
and the $H\rightarrow -H$ symmetries. In this regime, the density of dislocations
increases with increasing excess area $\Delta A_*$.
We also observe that dislocations are mostly found in pairs.
Within each pair, one dislocation can be deduced from the
other by the symmetry $H(X,Y)\rightarrow -H(X,Y)$.
In addition, the number of wrinkles passing between the two dislocations
varies from one pair to the other.

{\bf Regime IV.} For large shear velocity $V\geq V_\mathrm{max}$, dislocations disappear
and the membrane exhibits a perfect array of parallel stripes.
Some simulation movies for this case are shown in ESI\dag ({\it Movie 6}).
The final wavelength in this state, reported in Fig. \ref{fig:scalings_noshear}(b) with brown triangles
\footnote{The wavelength is evaluated from the simulation with the smallest value of the shear velocity $V$ that gives periodic state.}, 
is again in very good agreement
with the wavelength of the frozen state without shear
and with the sine ansatz.

Finally, note that, due to slowness in the numerical convergence
for large $\Delta A_*$ and large $V$, we have only explored systematically the cases $\Delta A_*<1$.
The few simulations performed with larger $\Delta A_*$ 
suggest a similar scenario.
Movies of these cases are reported in ESI\dag ({\it Movie 7}).

\subsection{Forces on the walls as a function of the shear rate $V$}
\label{sec:FmemV}

\begin{figure*}
\begin{center}
{\includegraphics[width=0.8\textwidth]{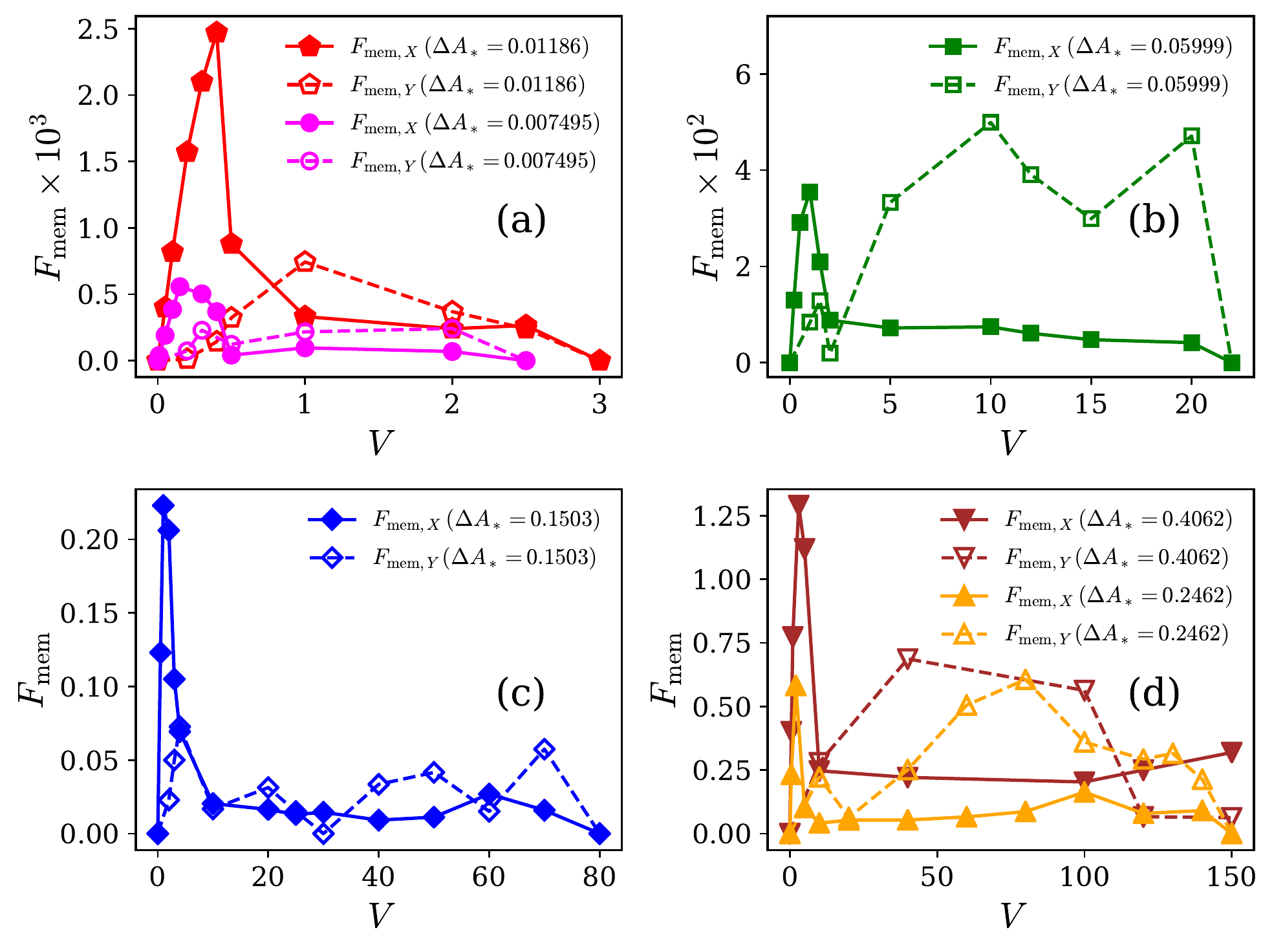}}
\caption{
Membrane  contribution to the tangential forces on each wall in steady state. $F_{\mathrm{mem}}$ is plotted
as a function of the shear rate $V$ for different excess areas $\Delta A_*$. 
Continuous lines: $F_{\mathrm{mem},X}$. Dashed lines: $F_{\mathrm{mem},Y}$.
The membrane excess area $\Delta A_*$ increases from Fig. (a) to (d). 
}
\label{fig:totalforce}
\end{center}
\end{figure*}

The 4 dynamical regimes of the membrane give rise to
distinctive behaviours of the tangential forces acting on the walls. 
The dependence of $F_{\mathrm{mem},X}$ on $V$ is summarized in the schematic at the bottom figure of Fig. \ref{fig:patternA0p1503}.
In Fig. \ref{fig:totalforce}, the forces measured in 
simulations are plotted as function of $V$ for different $\Delta A_*$.

In Regime I, the force $F_{\mathrm{mem},X}$ increases 
 linearly with the shear velocity $V$.
The force $F_{\mathrm{mem},Y}$
 is very  small and can exhibit both signs. 

\begin{figure}
\begin{center}
{\includegraphics[width=0.45\textwidth]{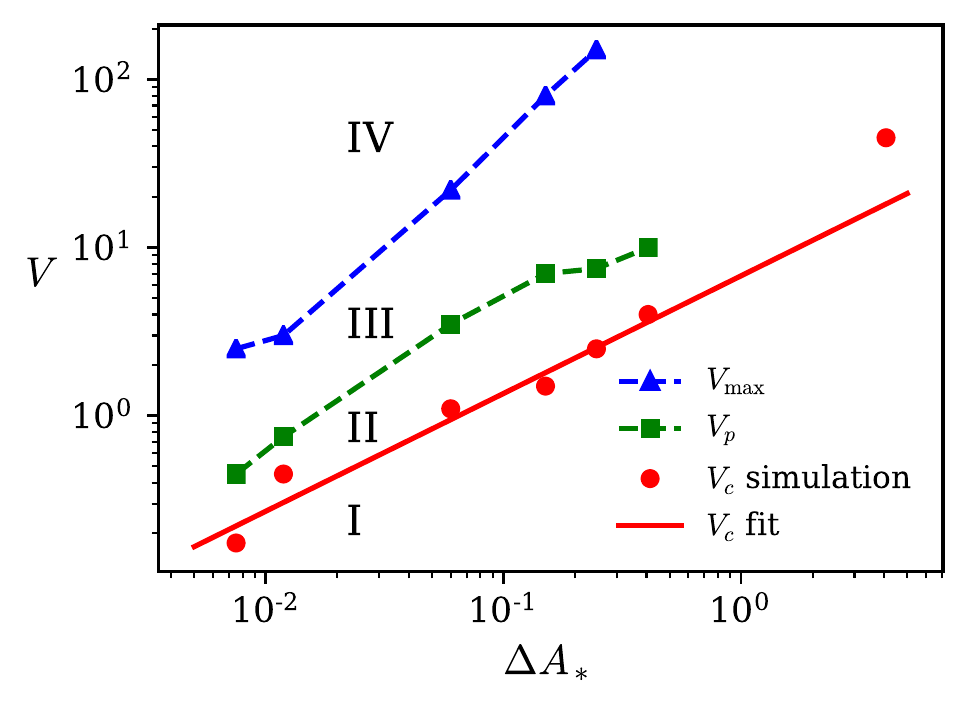}}
\caption{Non-equilibrium phase diagram with four different regimes.
These regimes correspond to those reported in Fig.~\ref{fig:patternA0p1503} and \ref{fig:totalforce}.
The three critical velocities $V_c$, $V_p$, and $V_\mathrm{max}$
are defined in the text.
The red solid line reports the fit $V_c=6.7941\Delta A_*^\frac{7}{10}$ with  the predicted 
exponent for $\Delta A_*\ll 1$, see Eq. (\ref{eq:V_c_of_p}).
Dashed lines are guide to the eye.
}
\label{fig:VcVmax}
\end{center}
\end{figure}

When the shear rate is increased further, the force increases slower than linearly and
at the critical shear rate $V_c$ the force reaches a peak.
The value of $V_c$ as a function of the excess area $\Delta A_*$ is reported in Fig. \ref{fig:VcVmax}.
By convention, and although some weak reorganisation
of the membrane pattern can be observed just before the peak,
we use $V>V_c$ as a formal definition of Regime II.
In Regime II, the steady-state force along $X$ drops quickly as $V$ increases.
When the dynamics is oscillatory, the forces along the $X$ and $Y$ directions are oscillatory, 
as shown in Fig. \ref{fig:oscillating_ripples}(b).

As the shear rate is increased further, the membrane
reaches Regime III dominated by dislocations where the force along $X$ presents a
noisy plateau. Simultaneously, the force along $Y$ becomes
very large in amplitude, but still with equal probability 
in the $+Y$ and $-Y$ directions.

Finally, in regime IV, as $V\geq V_\mathrm{max}$, the force vanishes
in both directions. 
The value of $V_\mathrm{max}$ as a function of the excess area is reported in Fig. \ref{fig:VcVmax}.

\section{Critical shear rate $V_c$}
\label{sec:Vcanalytic}

In this section, we obtain an estimate of the critical shear rate $V_c$
from a simple balance between different terms in
Eq. (\ref{eq:dynamic2Dnormalized}). In order to do so,
we compare the shear term 
$VH\partial_x H$ and terms associated to the force $F_Z$.
In the spirit of the sine-ansatz where the amplitude $a$ 
and wavelength $\lambda$ of the pattern result from the minimization
of the energy, the three terms in the force, associated to bending rigidity,
tension and confining potential,  have similar magnitudes.
We therefore choose arbitrarily to compare the shear term 
with the bending rigidity term, i.e. we assume that
$V_c$ is such that $VH\partial_XH\sim \Delta^2H$.

Assuming the sine-profile ansatz following the same lines as in the analyis of
the dynamics of the membrane without shear discussed above,
we set $H=a\cos(q\zeta)$ where $a$ is the amplitude, $q$ is the wave number and $\zeta$ 
is the space variable orthogonal to the wrinkles in the $X,Y$ plane. 
Each derivative brings a prefactor $q$ and
we obtain $C_0q^4a=V_cqa^2$, where $C_0$ is a number of the order of one. 
This relation is re-written as
\begin{equation}
V_c=C_0\frac{q^3}{a}.
\label{eq:Vc_qa}
\end{equation}

Since the membrane profile at small shear
rates is not affected by shear as observed in the simulations,
we assume that  $a$ and $q$ obey the same laws as in the 
case without shear. Thus, in the limit of small excess area,
we use Eq. (\ref{eq:general_analytical_noshear}), and obtain
\begin{equation}
\label{eq:V_c_of_p}
V_c=8C_0C_{m,p}^{-4} \Delta A_*^{\frac{3}{2}-\frac{8}{p+2}}.
\end{equation} 
Such a power-law behavior with $C_0 \approx 1.43$ is in good agreement with simulations at 
small $\Delta A_*$, as shown in  Fig. \ref{fig:VcVmax}.

In the limit of large excess area, the same procedure
(combining Eq. (\ref{eq:lambda_Alarge}) with Eq. (\ref{eq:Vc_qa})) leads to
\begin{align}
\label{eq:V_c_of_p_large_excess_area}
    V_c=8C_0\Delta A_*^{3/2},
\end{align} 
with a possibly different value of $C_0$ from that of the regime at small excess area. 
We could not obtain accurate simulations at large large $\Delta A_*$
and large $V$, and we can therefore not extract the 
value of $C_0$ for large $\Delta A_*$. 
However, we have obtained $V_c\approx 45$
for $\Delta A_*=4.07$, which suggests $C_0\approx 0.7$ from Eq. (\ref{eq:V_c_of_p_large_excess_area}).

\section{Forces in drifting steady-states}

In a drifting steady-states we can combine Eqs. (\ref{eq:memforce2Dnormalized}) and (\ref{eq:driftsteady_main}) 
to obtain useful expressions of the force exerted by the membrane on each wall
\begin{align}
\label{eq:memforce2Ddriftarrangednormalized}
&F_{\mathrm{mem},X}=12\left[
V\langle (H\partial_XH)^2\rangle
-V_{dX} \langle H(\partial_XH)^2\rangle
-V_{dY} \langle H\partial_XH\partial_YH\rangle
\right],\nonumber \\
&F_{\mathrm{mem},Y}=12
\left[ 
V\langle H^2\partial_XH\partial_YH\rangle
-V_{dX} \langle H \partial_XH\partial_YH\rangle
-V_{dY} \langle H(\partial_YH)^2\rangle
\right].
\end{align}
As discussed in the following, these expressions
allow one to study the forces on the basis 
of the analysis of the membrane profile.

\subsection{Small shear rates}
\label{sec:Vsmallresults}

For small shear rates in Regime I, the configuration of the membrane is
similar to that obtained in the absence of shear.
Since the configuration without shear is disordered and isotropic, 
the asymmetry of the membrane configuration
which breaks the $X\rightarrow-X$ or the $Y\rightarrow-Y$
symmetries in a large system is small and random.
As a consequence, the drift velocities $V_{dX}$ and $V_{dY}$ 
are also small and random.
Hence, the contributions proportional to these velocities in
Eqs. (\ref{eq:memforce2Ddriftarrangednormalized})
are negligible as compared to the first term proportional to $V$
which does not vanish for symmetric and isotropic configurations.

\begin{figure}
\begin{center}
{\includegraphics[width=0.4\textwidth]{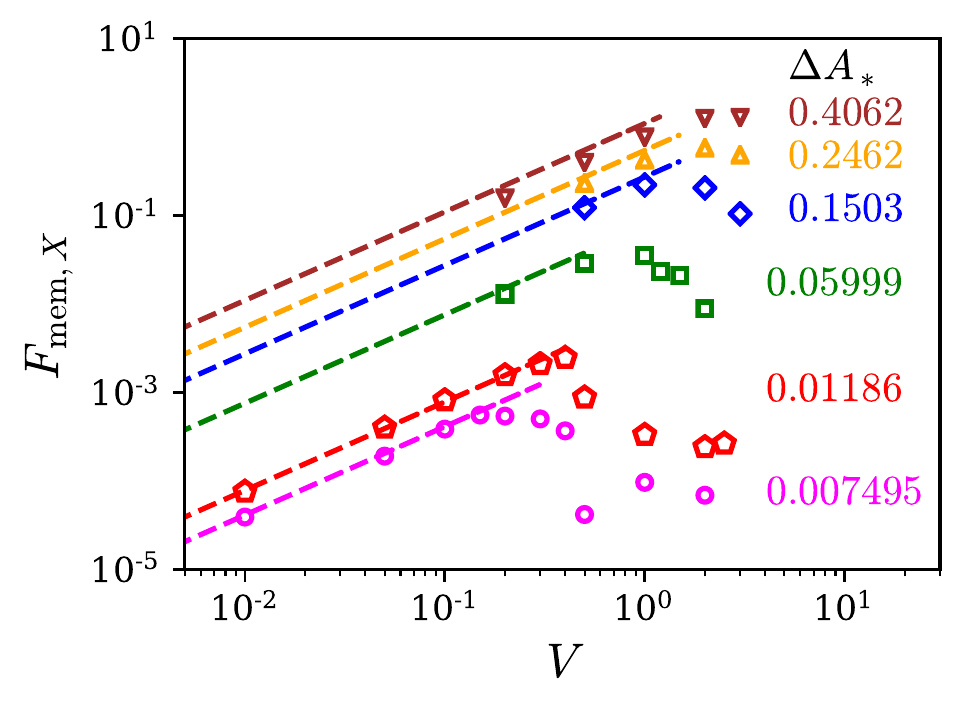}}
\caption{Regime I: linear behavior of the membrane force at small shear rates.
$F_{\mathrm{mem},X}$ in steady state
is plotted as a function of the shear rate $V$ for different $\Delta A_*$. 
The dashed lines are the linear
behavior predicted by Eq. (\ref{eq:H2dH2}).
Symbols: simulation results. The membrane excess area $\Delta A_*$ increases from the bottom curve to the top curve. 
}
\label{fig:fmem_Vmall}
\end{center}
\end{figure}

Therefore, to leading order, 
the membrane contribution to the force is then given by
\begin{equation}
\label{eq:memforce2D_Vsmall_normal}
F_{\mathrm{mem},X}=12V \langle (H\partial_XH)^2 \rangle.
\end{equation}
The spatially averaged quantity $\langle (H\partial_XH)^2 \rangle$ is
evaluated using the sine-profile ansatz.
Since the labyrinthine pattern is isotropic,
we average over all possible orientations of the wrinkles. 
We therefore define the angle $\theta$ between the wrinkle orientation
and the $Y$ axis. In the sine ansatz Eq. (\ref{eq:sineansatz_def}), the coordinate 
orthogonal to the wrinkle then reads $\zeta=X\cos\theta+Y\sin\theta$.
The average takes the form
\begin{equation}
\langle(H\partial_XH)^2  \rangle=\frac{1}{2\pi}\int_0^{2\pi}d\theta \frac{1}{\lambda}\int_0^{\lambda} d\zeta(H\partial_XH)^2 ,
\end{equation}
where $H=a\cos(q\zeta)$ and $\partial_XH=-aq\cos\theta\sin(q\zeta)$. Performing the integration on $\zeta$ and $\theta$,
we obtain
\begin{equation}
\label{eq:H2dH2_general}
\langle (H\partial_XH)^2 \rangle=\frac{a^4q^2}{16}.
\end{equation}
Since the labyrinthine pattern is not affected by shear for small $V$, 
we substitute the expressions of $a$ and $q$ that are obtained from the energy minimization
of the sine-ansatz without shear in Eq. (\ref{eq:H2dH2_general}).
Thus, in regime I, i.e., for small $V$ we obtain from Eq. (\ref{eq:memforce2D_Vsmall_normal})
\begin{align}
    F_{\mathrm{mem},X}=\frac{3}{4}Va^4q^2.
\end{align}
In the limit of small $\Delta A_*$, this leads to
\begin{align}
\label{eq:H2dH2}
    F_{\mathrm{mem},X}=3 C_{m,p}^2 V \Delta A_*^{1+\frac{4}{p+2}},
\end{align}
where $C_{m,p}$ is given in Eq. (\ref{eq:Cmp}).
This expression is in quantitative agreement with the simulation results,
as shown in Fig. \ref{fig:fmem_Vmall}.
In the limit of large excess area, we obtain
\begin{align}
\label{eq:H2dH2_large_excess_area}
    F_{\mathrm{mem},X}=3V \Delta A_*.
\end{align}

\subsection{Large shear rates: defect dependent membrane forces}

\begin{figure*}
\begin{center}
    {\includegraphics[width=1\textwidth]{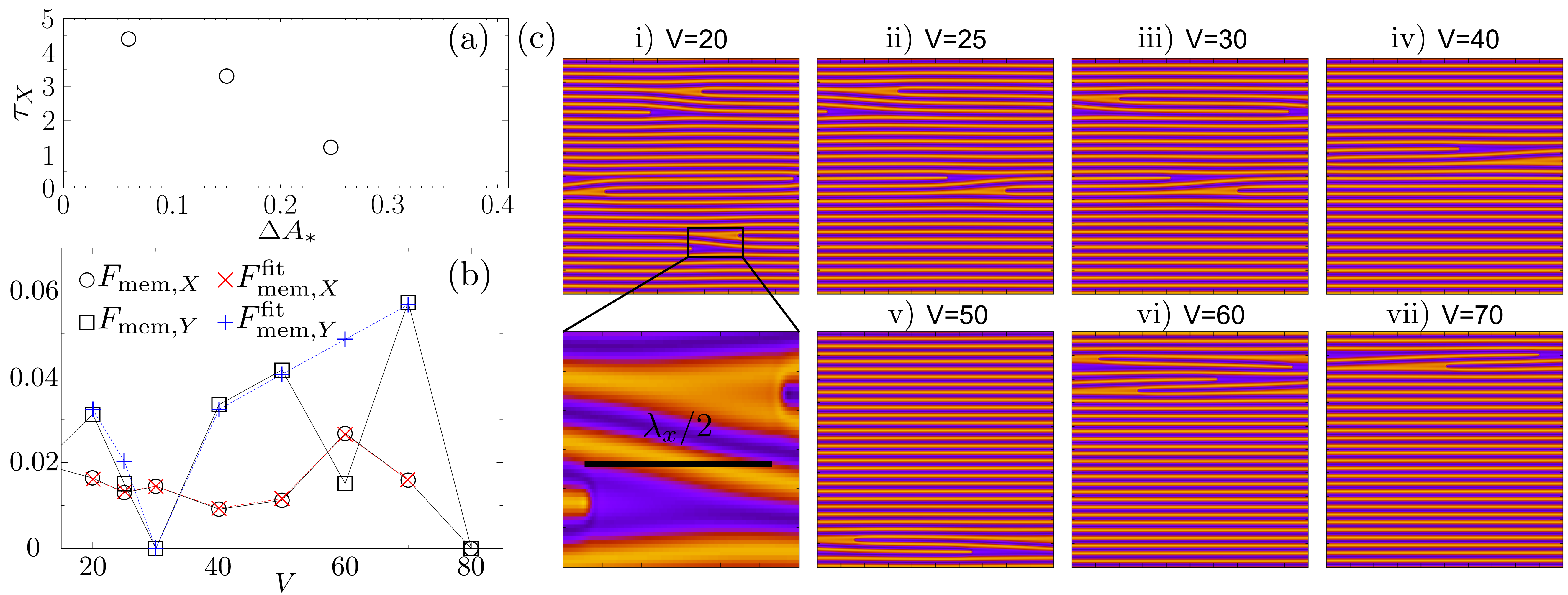}}
\caption{
Regime III: dislocation-dominated forces.
(a) Evolution of $\tau_x$ from Eq. (\ref{eq:lambdaX_tauX_V}) with excess area $\Delta A_*$. 
(b) The black solid line with $\bigcirc$ (resp. $\Box$) represents the membrane force $F_{\mathrm{mem},X}$ (resp. $F_{\mathrm{mem},Y}$) obtained in simulations for $\Delta A_*=0.1503$ and evaluated from Eqs. (\ref{eq:memforce2Dnormalized}). 
The red dotted line with {\color{red}$\times$} is a multiple linear regression of $F_{\mathrm{mem},X}$ for $V\geq20$ using Eq. (\ref{eq:forcememxapprox})
($a_{\mathrm{wr,X}}\approx 32.4$ and $a_{\mathrm{dis,X}}\approx 8.6$). 
The blue dotted line with {\color{blue}$+$} is a linear regression of $F_{\mathrm{mem},Y}$ using Eq. (\ref{eq:forcememyapprox}) 
($a_{\mathrm{wr,Y}}=35.4$, 
the simulations $V=30$ and $V=60$ have been removed for linear regression, see text). 
(c) Membrane profile in steady-states corresponding to the points in panel (b).
Values of $(n,N,n_0)$ (see text for definition) extracted from the images for various shear rates: 
i): $V=20$ (4,3,2), ii): $V=25$ (3,2,1), iii): $V=30$ (2,2,0) iv): $V=40$ (1,1,1), v): $V=50$ (1,1,1), vi): $V=60$ (1,2,1) and vii): $V=70$ (1,1,1). 
The zoom shows that a half-wavelength $\lambda_X/2$ in the tilted wrinkles zone corresponds approximately to
the distance between the two dislocation cores in a pair.
}
\label{fig:dislocations}
\end{center}
\end{figure*}

For large shear rates in regime III, the configuration of the membrane 
is composed of a few dislocations on a set of almost parallel
stripes aligned along $X$. Thus, the configuration
is globally anisotropic. The background of parallel
wrinkles along $X$ provides no contribution to the force exerted by the 
membrane on the walls from Eqs. (\ref{eq:memforce2Ddriftarrangednormalized}).
The dislocation pairs therefore can be seen as elementary building blocks,
each pair providing its own contribution to the total force.

A close inspection of the images of the membrane in Fig. \ref{fig:dislocations}
indicates that for each pair of dislocations, tilted wrinkles pass between the two dislocations. 
Each wrinkle  is approximately
shifted by one period along $Y$ when passing between the two dislocation of a dislocation pair.
We therefore design a simple approximation where all the wrinkles
passing between the pairs are straight and identical, and are tilted
only in a zone of length $\lambda_X$ along $X$.
This leads to the sine-profile ansatz 
\begin{equation}
\label{eq:tilted_sine_ansatz}
H=a\cos\left(\vartheta q_XX+q_YY\right),
\end{equation}
with $\vartheta=1$ or $-1$ when the wrinkle slope is negative or positive respectively, 
$q_X=2\pi/\lambda_X$ where $\lambda_X$ is the wavelength of tilted wrinkles along $X$-axis or shear direction (see Fig. \ref{fig:dislocations}). The quantities $q_Y$ and $\lambda_Y$ are defined in a similar fashion.

The membrane contributions to the force Eq. (\ref{eq:memforce2Ddriftarrangednormalized})
is integrated over the size $\lambda_X$ of the tilted zone
 along $X$. In order to isolate the contribution
of each tilted wrinkle, we also integrate  over a single period $\lambda_y$  along $Y$.
Using the ansatz in Eq. (\ref{eq:tilted_sine_ansatz}),
the contributions proportional to $V_{dX}$ and $V_{dY}$ vanish, and we find
that one single period of tilted wrinkle along $X$ and along $Y$ exerts the forces
\begin{align}
\label{eq:F_Wr_memX}
&F_{\mathrm{mem},X}^{\mathrm{wr}}=\frac{1}{L_XL_Y}6\pi^2V\frac{\lambda_Y}{\lambda_X}a^4, \\
\label{eq:F_Wr_memY}
&F_{\mathrm{mem},Y}^{\mathrm{wr}}=\frac{1}{L_XL_Y}6\pi^2Va^4\vartheta.
\end{align}

Moreover, simulations
in regime III are consistent with a linear increase
of $\lambda_X$ with the shear rate:
\begin{align}
\label{eq:lambdaX_tauX_V}
\lambda_X=\tau_XV 
\end{align}
with $\tau_X$ a constant.
A measurement of $\lambda_X$ based on the distance between two successive 
zeros of the membrane profile in the tilted region allows one to estimate $\tau_X$ by linear regression. 
The value of $\tau_X$ as a function of  $\Delta A_*$ is reported in Fig. \ref{fig:dislocations} (a).
In addition, since the angle of the wrinkles with the $X$ direction is small, we simply assume
that $\lambda_Y\simeq\lambda$. Combining these assumptions for $\lambda_X$ and $\lambda_Y$ 
Eq. (\ref{eq:F_Wr_memX}) is written as
\begin{align}
\label{eq:F_Wr_memX_2}
&F_{\mathrm{mem},X}^{\mathrm{wr}}=\frac{1}{L_XL_Y}12\pi^3V\frac{a^4}{q\tau_X}.
\end{align}

In addition to the contribution of the tilted
wrinkles, there is also a contribution to the force caused by the cores of the dislocations.
As discussed above, the second dislocation in a pair is obtained by a symmetry 
$H(X,Y)\rightarrow -H(-X,Y)$ from the first one. 
Moreover an analysis of the simulation images reveals the
approximate symmetry of each dislocation core under the transformation $Y\rightarrow-Y$. 
Using these symmetry properties and 
Eqs. (\ref{eq:memforce2Ddriftarrangednormalized}),
we obtain the contributions for one pair of dislocation cores
\begin{equation}
\label{eq:force_dislocation_pair}
F_{\mathrm{mem},X}^{\mathrm{disloc,pair}}=24V\langle (H\partial_XH)^2 \rangle_{core}, \quad F_{\mathrm{mem},Y}^{\mathrm{disloc,pair}}=0,
\end{equation}
where $\langle (H\partial_XH)^2 \rangle_{core}$ is evaluated
by means of integration around one single dislocation core
\begin{equation}
\langle (H\partial_XH)^2 \rangle_{core}=\frac{1}{L_XL_Y} \int_{core}dX \int_{core}dY (H\partial_XH)^2.
\end{equation}

Assuming that the profile of the dislocation core does not depend
on $V$, the contributions to the membrane forces take the form
\begin{equation}
F_{\mathrm{mem},X}^{\mathrm{wr}}=\frac{a_{\mathrm{wr},X}}{L_XL_Y}, \quad
F_{\mathrm{mem},Y}^{\mathrm{wr}}=\vartheta\frac{a_{\mathrm{wr},Y}}{L_XL_Y}V, \quad
F_{\mathrm{mem},X}^{\mathrm{disloc,pair}}=\frac{a_{\mathrm{dis},X}}{L_XL_Y}V,
\end{equation}
where $a_{\mathrm{wr},X}$, $a_{\mathrm{wr},Y}$ and  $a_{\mathrm{dis},X}$
are constants independent of $V$, $L_X$, $L_Y$, and $\vartheta$
\begin{align}
\label{eq:a_wrX}
&a_{\mathrm{wr},X}=12\pi^3 \frac{a^4}{\tau_Xq},  \\
\label{eq:a_wrY}
&a_{\mathrm{wr},Y}=6\pi^2 a^4.
\end{align}
The third constant, which accounts for the contribution of the dislocation core
is unknown
\begin{align}
\label{eq:a_disX_p8_m1}
&a_{\mathrm{dis},X}=24\int_{core}dX \int_{core}dY (H\partial_XH)^2.
\end{align}

Summing the contributions of all dislocation pairs in the system, the total force exerted on each wall can be written as
\begin{align}
\label{eq:forcememapproxtot}
&F_{\mathrm{mem},X}= \frac{n}{L_XL_Y} F_{\mathrm{mem},X}^{\mathrm{wr}} + \frac{N}{L_XL_Y} F_{\mathrm{mem},X}^{\mathrm{disloc,pair}},\nonumber \\
&F_{\mathrm{mem},Y}= \frac{n_0}{L_XL_Y} |F_{\mathrm{mem},Y}^{\mathrm{wr}}|,
\end{align}
where  $n$ is 
the total number of wrinkles within pairs of dislocations, 
$n_0$ is the number of wrinkles with negative slope minus the number of wrinkles with positive slope 
and $N$ is the total number of pairs of dislocations.
The total forces Eqs. (\ref{eq:forcememapproxtot}) may be rewritten as
\begin{align}
&F_{\mathrm{mem},X}=a_{\mathrm{wr},X}\frac{n}{L_XL_Y}+a_{\mathrm{dis},X}\frac{N}{L_XL_Y}V, \label{eq:forcememxapprox}\\
&F_{\mathrm{mem},Y}=a_{\mathrm{wr},Y}\frac{n_0}{L_XL_Y}V\label{eq:forcememyapprox}.
\end{align}
As seen from Fig. \ref{fig:dislocations} (b),
the Eqs. (\ref{eq:forcememxapprox},\ref{eq:forcememyapprox})
provides an excellent fit to the simulation results.
For $\Delta A_*=0.1503$ a linear regression leads to 
$a_{\mathrm{wr},X}^{\mathrm{fit}}\approx32.4$, 
$a_{\mathrm{dis},X}^{\mathrm{fit}}\approx8.6$ 
and $a_{\mathrm{wr},Y}^{\mathrm{fit}}\approx35.4$.

However, two remarks are in order. First,
the fit fails to describe the rare cases when two dislocation
pairs are very close to each other.
This is for example the case in  Fig. \ref{fig:dislocations} (b) and (c)(vi)
for $V=60$.
Second, the analytical expressions 
of the constants Eqs. (\ref{eq:a_wrX},\ref{eq:a_wrY})
provide the correct order of magnitude
but are not quantitatively accurate. Indeed, 
using the expressions of $a$ and $q$ from the
sine ansatz in the limit of small excess area Eq. (\ref{eq:general_analytical_noshear}), they lead to
$a_{\mathrm{wr},X}\approx41.6$
and $a_{\mathrm{wr},Y}\approx21.6$
for  $\Delta A_*=0.1503$.
These discrepancies could originate in the 
crudeness of the assumptions leading to Eqs. (\ref{eq:a_wrX},\ref{eq:a_wrY}),
such as straight tilted wrinkles, and the absence of smooth decay
of the perturbation away from the dislocation pair.
Despite the difficulty in finding an accurate
analytical expression for the prefactors $a_{wr}$ in Eqs. (\ref{eq:forcememxapprox},\ref{eq:forcememyapprox}), 
the simple dependence of the forces on the topological numbers $n$, $N$ and $n_0$
is remarkable.

Since the asymmetry of the membrane profile in steady-state emerges from the randomness of initial conditions,
we expect that $n_0/n\rightarrow 0$ and $n_0/N\rightarrow 0$ as $L_X,L_Y\rightarrow \infty$,
in the limit of large systems.
Therefore, the dominant contribution to the force would be along $X$. However, our simulations are performed in a finite simulation box 
and the number of dislocations usually does not exceed 10. Hence at large shear rates when the terms proportional
to $V$ dominate, since the prefactor of the force due to tilted wrinkles
is larger in the $Y$ direction $a_{\mathrm{wr},Y}>a_{\mathrm{dis},X}$, 
we often observe that 
$F_{\mathrm{mem},Y}=a_{\mathrm{wr},Y}n_0V>F_{\mathrm{mem},X}\approx a_{\mathrm{dis},X}NV$
although $n_0\leq N$. However, the forces along $X$ should dominate
in the limit of very large systems where we expect $N\gg n_0$.

\section{Critical shear rate $V_\mathrm{max}$}

The transition to regime IV, where all dislocations
disappear leading to a perfect array of parallel wrinkles along $X$, 
occurs at large shear rates. Here, we present evidences
that this transition could originate in a finite size effect.

Indeed, when the shear rate is very large,
the distance between the dislocations within a pair increases (see Eq. (\ref{eq:lambdaX_tauX_V}))
and becomes comparable to the system size.
The distance between two dislocation cores within a pair
is roughly equal to $\lambda_X/2$, as seen from the zoom in Fig. \ref{fig:dislocations}(c).
Due to the periodic boundary conditions in the $X$ direction,
the complementary distance between the two dislocations is $L_X-\lambda/2$.
The transition roughly occurs when these two distances are similar $\lambda_X/2\approx L_X-\lambda_X/2$,
i.e., when $\lambda_X\approx L_X$.
Using Eq. (\ref{eq:lambdaX_tauX_V}), this criterion leads to $\tau_XV/L_X\approx 1$.

\begin{figure}
\begin{center}
    {\includegraphics[width=0.4\textwidth]{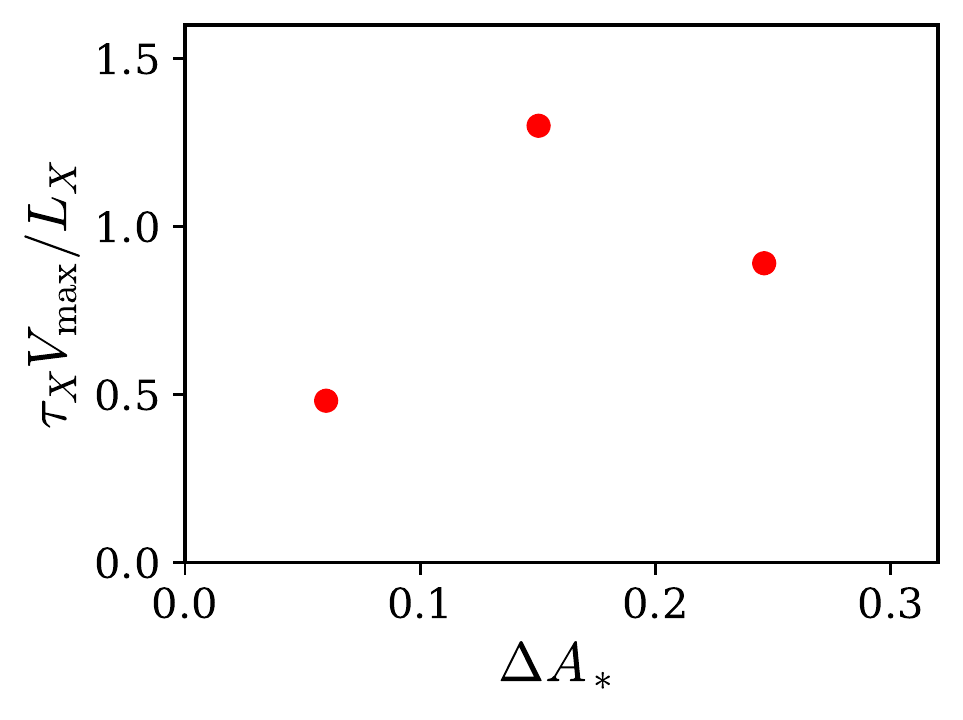}}
\caption{
Finite size effects in the transition to Regme IV.
The ratio $\tau_X V_\mathrm{max}/L_X$ is plotted for three excess areas $\Delta A_*$ with $L_X=200$.
The values of $\tau_X$ are from Fig. \ref{fig:dislocations}(a).
}
\label{fig:tauXVmaxLX}
\end{center}
\end{figure}

In Fig. \ref{fig:tauXVmaxLX}, we show the ratio $\tau_XV/L_X$
for three values of $\tau_X$ that have been
extracted from simulations in Regime III (see  Fig. \ref{fig:dislocations}(a)).
Fluctuations are large in Fig. \ref{fig:tauXVmaxLX} due to poor statistics
with few simulations and few dislocations in each simulation.
However, the ratio $\tau_XV/L_X$ is around one in all cases, 
and these results therefore support the hypothesis of a transition
to regime IV controlled by finite size effects.

\section{Summary and Discussions}
\subsection{Summary of results}
\label{sec:sumresul}

In summary, we have studied the dynamics of a lipid membrane
confined between two flat walls in the presence and in the absence of shear. 
We have also evaluated the tangential forces exerted by the membrane on each wall
due to its complex nonlinear dynamics in the presence of shear. In this section,
we provide a concise recapitulation of the main results.

In the absence of shear, the membrane forms a disordered labyrinthine pattern of wrinkles.
The wavelength and amplitude of the pattern are described quantitatively using a sine-ansatz.
The results depend on 
the free energy potential that confines the membrane between the walls. 
We have used the confinement potential
\begin{align}
{\cal U}(h)= \frac{{\cal U}_0}{[1-(h/h_0)^p]^m}.
\end{align}
Two different regimes are identified, depending on the
normalized excess area
\begin{align}
\label{eq:physical_excess_area}
\Delta A_*=\frac{\Delta {\cal A}}{{\cal L}_x{\cal L}_y}\,\frac{\ell_\parallel^2}{h_0^2},
\end{align}
where $\ell_\parallel$ is defined in Eq. (\ref{eq:normalisationdistance}).
In physical units, we have
\begin{align}
\lambda&= 
 \pi \ell_\parallel\times
\left\{ \begin{array}{ll}
    C_{m,p} \Delta A_*^{-\frac{1}{2}+\frac{2}{p+2}} &  \Delta A_*\ll 1,\\
    \Delta A_*^{-\frac{1}{2}} &  \Delta A_*\gg 1,\\
\end{array} \right.
\\
    \langle h^2 \rangle^{1/2}&=  
    2^{-\frac{1}{2}}h_0 \times
\left\{ \begin{array}{ll}
     C_{m,p}  \Delta A_*^\frac{2}{p+2} &  \Delta A_*\ll 1,\\
    1  &  \Delta A_*\gg 1,
\end{array} \right.
\end{align}
where $C_{m,p}$ is a number provided in 
Eq. (\ref{eq:Cmp}). Here we use the standard deviation of the height
$\langle h^2 \rangle^{1/2}$ which is defined for any 
observable membrane profile, rather than the amplitude $a$
which is well defined in the case of the sine ansatz only.
They are related via $\langle h^2 \rangle^{1/2}=2^{-1/2}a$. Moreover,
note that the limit $\Delta A_*\gg 1$
can be obtained from the limit $p\rightarrow\infty$ in the regime of small normalized
excess area $\Delta A_*\ll 1$.
Our results obtained with a single-well potential are in contrast with the complex 
behavior observed previously with a double-well adhesion potential\cite{membraneadhesion2D2018},
where endless coarsening dynamics was found for low normalized excess area.

In the presence of shear, the dynamics
exhibits four different regimes
depending on the shear rate $\upsilon_0$.
In these regimes, shear does not affect the wavelength
of the wrinkles, but is able to reorganise the wrinkles
when the shear rate is large enough.

{\bf Regime I} is found at small shear rates. The labyrinthine pattern is then essentially unaffected by 
shear. The tangential force exerted by the membrane on each wall per unit
area of wall is
from Eq. (\ref{eq:H2dH2})
\begin{align}
f_{mem,x} &=
\upsilon_0\frac{\mathcal{U}_0^\frac{1}{2}h_0}{4\nu\kappa^\frac{1}{2}} \times
\left\{ \begin{array}{ll} 
    C_{m,p}^2 \Delta A_*^{1+\frac{4}{p+2}} &  \Delta A_*\ll 1,\\
    \Delta A_*  &  \Delta A_*\gg 1.
\end{array} \right.
\end{align}
The forces along $y$ are small in this regime.

We found that there is a critical shear rate 
above which the wrinkle pattern starts to reorganize
\begin{align}
\frac{\upsilon_{c}}{h_0}&= \frac{4\nu {\cal U}_0^\frac{3}{4} \kappa^\frac{1}{4}}{ h_0^\frac{5}{2}} \times
\left\{ \begin{array}{ll} 
    C_0C_{m,p}^{-4} \Delta A_*^{\frac{3}{2}-\frac{8}{p+2}}&  \Delta A_*\ll 1,\\
    C_0\Delta A_*^\frac{3}{2}  &  \Delta A_*\gg 1,
\end{array} \right.
\end{align}
where $C_0$ is a number of the order of one that depends on $m$ and $p$
and is different in the regimes of small and large normalized excess area
(we found $C_0\approx 1.43$ for $m=1$ and $p=8$ in the regime $\Delta A_*\ll 1$). 

For shear rates larger than $\upsilon_{c}$, wrinkles have 
the tendency to align along the shear direction.
As a consequence, the membrane contribution
to the forces on each wall drops. This is {\bf regime II}.
In this regime, we rarely observe oscillatory membrane
configurations that give rise to oscillatory forces on the walls.

When the shear rate is increased further, we reach {\bf regime III}
where the wrinkles are mainly aligned along the shear direction, 
with some localized dislocation defects
that are grouped in pairs. The resulting
contribution of the membrane to the forces along $x$ depends linearly on the number $N$ 
of dislocation pairs and on the total number $n$ of wrinkles
passing between the dislocations within dislocation pairs.
The force per unit wall area reads
\begin{align}
f_{\mathrm{mem},x}&= 
 \frac{N}{{\cal L}_x {\cal L}_y}  
\frac{2\upsilon_0}{\nu}  
\left(\int\int_{core} \mathrm{d}x \,\mathrm{d}y\, (h\partial_xh)^2 \right)
\nonumber \\
&+
\frac{n}{{\cal L}_x {\cal L}_y} 
\frac{\pi^3 }{4\tau_X} \kappa^\frac{1}{4} h_0^\frac{1}{2}{\cal U}_0^\frac{3}{4}  \times
\left\{ \begin{array}{ll} 
    C_{m,p}^5\Delta A_*^{-\frac{1}{2}+\frac{10}{p+2}}&  \Delta A_*\ll 1,\\
    \Delta A_*^{-\frac{1}{2}}  &  \Delta A_*\gg 1.
 \end{array} \right.
\end{align}
The contribution to the forces along $y$ is proportional
to the difference $n_0$ between the number of negative-slope wrinkles and positive-slope wrinkles
within dislocation pairs
\begin{align}
f_{\mathrm{mem},y}&= 
\frac{n_0}{{\cal L}_x {\cal L}_y} 
\frac{\pi^2 }{2}  \frac{\upsilon_0h_0^2}{\nu}   \times
\left\{ \begin{array}{ll} 
C_{m,p}^4\Delta A_*^\frac{8}{p+2}&  \Delta A_*\ll 1,\\
    1  &  \Delta A_*\gg 1.
 \end{array} \right.
\end{align}
The dimensionless prefactor of the transverse forces along $y$ is found to be rather large in this regime.
As a consequence, significant transverse forces could arise 
in physical systems with a finite size due to unbalanced
statistical fluctuations leading to a non-vanishing $n_0$.

Finally, at very large shear velocities above a threshold value $\upsilon_0>\upsilon_\mathrm{max}$,
we find {\bf regime IV} where
dislocations disappear and the membrane profile is composed of a perfectly ordered set
of wrinkles aligned along the shear direction.
In this regime, the membrane exerts no tangential force on the walls.

One of the limitations of our numerical investigations is the fact that the simulation
boxes were finite. Due to this limitation, we could not reach the asymptotic
limit of large systems where lateral forces along $y$ are expected
to be negligible as compared to forces along $x$ in Regime III.
In addition, the transition to regime IV at $V=V_\mathrm{max}$ could be
controlled by the system size.  However, experimental
systems also exhibit a finite size, and the finite-size effects 
observed in our simulations could also be relevant to some experimental conditions.

Another limitation of our study is the lack of 
systematic numerical investigation of the regime at very large
normalized excess area and large shear rates. As a consequence, the analytical
predictions at large normalized excess area have not been all checked
in details numerically. However, two results suggest
that these predictions should be valid.
First, simulations at $\Delta A_*\approx 4$
provide similar results as those observed for small $\Delta A_*$.
In addition, the expressions in the regime of large excess area
can be retrieved by taking the limit $p\rightarrow\infty$
in the expressions of the regime at small excess area 
which have been checked quantitatively in simulations.
We are therefore confident that our predictions for
large $\Delta A_*$ should apply quantitatively.

\subsection{Global rheological behaviour of the system}

\begin{figure*}
\begin{center}
    {\includegraphics[width=\textwidth]{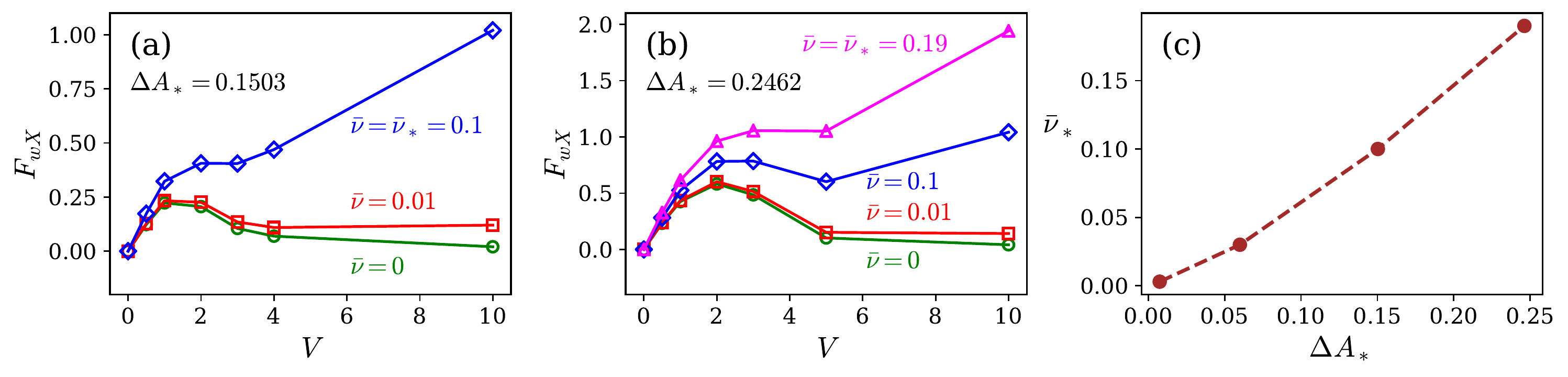}}
\caption{Non-monotonicity of the total tangential force.
Figs. (a) and (b) show the total normalized force $F_{wX}$ as a function of $V$ for different values of $\bar{\nu}$
.
(a) $\Delta A_*=0.1503$, (b) $\Delta A_*=0.2462$.
Fig. (c) shows the estimated value $\bar{\nu}_*$ 
(above which the total force is monotonic), as a function of $\Delta A_*$.
}
\label{fig:figtotalforce}
\end{center}
\end{figure*}

The total force is the sum of the membrane contribution 
with the viscous force of the liquid which is proportional to the shear rate.
Combining Eqs. (\ref{eq:totalforce2D},\ref{eq:normalizedtotalforce}), we have 
\begin{equation}
\label{eq:totalforce2D_beta}
  f_{wx}
  =\frac{\mathcal{U}_0^\frac{5}{4}}{24\kappa^\frac{1}{4}h_0^\frac{1}{2}}\left[\bar{\nu} V+12\langle H\partial_XHF_Z \rangle \right].
\end{equation}
The dimensionless parameter $\bar{\nu}=12\kappa^{1/2}\mu\nu/(\mathcal{U}_0^{1/2}h_0^2)$ 
governs the balance between the membrane contribution and the bulk liquid viscous force. 
Small $\bar{\nu}$ means that the membrane contributes significantly to the global friction while large $\bar{\nu}$ means that friction is dominated by viscous
dissipation in the liquid.
However, as discussed in Ref.\cite{membraneadhesion2D2018}, the parameter $\bar{\nu}$ also accounts for the
distinction between the limit of permeable walls for $\bar{\nu} \gg 1$ where our dynamical model Eq. (\ref{eq:dynamic2D}) apply,
and the limit of impermeable walls $\bar{\nu} \ll 1$ where instead, a different model accounting
for a fluid flow which is mainly parallel to the membrane, must be used. This latter conserved
model has been described in Ref.\cite{membraneadhesion2D2018}. In the present work,
we focus on the limit of permeable walls so that $\bar{\nu} \gg 1$.

As reported above, the membrane contribution to the force on the walls 
exhibits a non monotonic variation as a function of the shear rate. 
Indeed, it decreases in regime II. 
Using the normalized equation (\ref{eq:normalizedtotalforce}), 
we see that the total force can also be decreasing in regime II if  
$\bar{\nu}<-12\partial_V\langle H\partial_XHF_z \rangle$.
As seen in Fig. \ref{fig:figtotalforce}(a), such a condition
can for example be achieved when $\bar{\nu}<10^{-1}$ for $\Delta A_*=0.1503$.
We define the value $\bar{\nu}_*$ of $\bar{\nu}$ for which the total force changes from non-monotonic to monotonic variation. 
Fig. \ref{fig:figtotalforce}(c) shows that $\bar{\nu}_*$ increases almost linearly with $\Delta A_*$.

Since $\bar{\nu} \gg 1$ in the case of permeable walls, a non-monotonic 
dependence of the total force on the shear rate cannot be achieved
for small normalized excess area. We do not have detailed data
at large excess area. However, an extrapolation of 
the value of $\bar{\nu}_*$ to larger excess area suggests that
$\bar{\nu}_*>1$ roughly when  $\Delta A_*>1$. As a consequence, 
a non-monotonic behavior could be obtained at larger excess area.
Such a non-monotonic behavior was already suggested in the 1D model of Ref.\cite{PierreLouis2017}
in the impermeable limit, and could lead to instabilities such as stick-slip.

Let us now discuss quickly the orders of magnitude of relevant physical parameters.
The typical bending rigidity of lipid membrane is $\kappa\sim10^{-19}~\mathrm{J}$\cite{Henriksen2006, Pakkanen2011}. 
Moreover, several experiments \cite{Bruinsma2000, Weikl2009, Sengupta2010} suggest 
that the typical distance $h_0$ with the walls is a few nanometers in biological systems.
We also consider that the fluid surrounding the membrane is water with a viscosity $\mu\sim10^{-3}~\mathrm{Pa.s}$. 
For walls such as collagen \cite{Mironmendoza2010} or the cytoskeleton, 
the order of magnitude of the permeability can be evaluated as $\nu\sim~10^{-5}~\mathrm{m}^2.\mathrm{s}.\mathrm{kg}^{-1}$ 
using Darcy's law. 
The order of magnitude of the potential varies also strongly with the nature of the interaction.
In order to determine $\mathcal{U}_0$, we use the difference $\Delta \mathcal{U}=\mathcal{U}(h=0.9h_0)-\mathcal{U}(h=0)$
of potential between the potential in the center at $h=0$, and close to the walls at $h=0.9h_0$.
In the case of hydration forces we obtain 
$\mathcal{U}_0\sim10^{-2}-10^{-1}~\mathrm{J}.\mathrm{m}^{-2}$ \cite{Swain2001,Schneck2012}.
These numbers suggest $\bar{\nu}\sim 10$, and are therefore consistent with the 
case of permeable walls. In addition, this leads to $\ell_\parallel\sim h_0$ and $\epsilon\sim 1$.
This means that the units of the wrinkle wavelength $\lambda$ in Fig. \ref{fig:scalings_noshear}(b)
is about $10$nm. 
Furthermore, an inspection of Fig. \ref{fig:scalings_noshear}(b)
indicates that the small slope approximation, where $\lambda\gg h_0$
should be valid only when $\Delta A_*\ll 1$.
Hence, the excess area in physical coordinates 
$\Delta {\cal A}/({\cal L}_x{\cal L}_y)=\epsilon \Delta A_*\sim \Delta A_*$
should be small.
Although the fact that $\epsilon\sim 1$ questions the strict validity
of the small slope limit for larger excess areas,
our model should grasp some features of the dynamics of lubricated
contacts with membranes for small 
excess areas.

\subsection{Conclusion}

As a summary, we have presented a lubrication model
which describes the dynamics of an inextensible membrane
in an incompressible fluid between two walls. In quiescent conditions, the membrane forms
a frozen labyrinthine pattern of wrinkles which stores the membrane excess area.
When shear flow is induced by the 
parallel motion of the walls, the wrinkles reorganize
leading to a nonlinear rheological
response of the system. 

Our results show how the excess
area of the membrane participates in a non-trivial way 
to the rheology of contacts containing lipid membranes.
One important difference as compared to our previous investigations
of confined membranes is that we have considered confinement
by a single-well potential here as compared to the double-well
confinement potential used in our previous studies~\cite{PierreLouis2017,membraneadhesion2D2018}.
As a consequence of this difference, we observe no coarsening here,
while coarsening in the presence of a double-well potential 
was shown to be at the origin of thixotropic behavior in Ref.~\cite{PierreLouis2017}.
These results show that the rheological response of lubricated contacts
containing membranes could be controlled by the details of the confinement potential.

Finally, although our simple model catches
a complex behavior from a set of simple physical ingredients,
much yet remains to be done to describe the role of 
membranes in specific biolubrication systems.
In particular, possible future extensions of our work include the cases of impermeable walls
and of stacks of several membranes.
In addition, one could study the 
role of thermal fluctuations, e.g. following a Langevin approach as in Ref.~\cite{LeGoff2014}.

\section*{Author contributions}
TLG and TBTT have contributed equally
to the manuscript.
Conceptualization: OPL.
Formal analysis and Writing: TLG, TBTT and OPL.
Software: TBTT and TLG.

\section*{Conflicts of interest}
There are no conflicts to declare.

\section*{Acknowledgements}
The authors acknowledge support from Biolub Grant No. ANR-12-BS04-0008.
TBTT acknowledges support from CAPES (grant number PNPD20130933-31003010002P7) and FAPERJ (grant number E-26/210.354/2018). 
We thank Prof. F. D. A. Aar\~ao Reis for organizing the CAPES-PrInt program at the Universidade Federal Fluminense in December 2019 when part of the project was carried out.

\begin{appendix}

\section{Derivation of the force on the walls}
\label{secapp:lubric_force}

\subsection{Hydrodynamic flow}
In the lubrication approximation, 
the components of the fluid velocity $\upsilon_x,\upsilon_y$
take the form of a Poiseuille flow \cite{membraneadhesion2D2018}
\begin{equation}
\label{aeq:poiseuille}
\upsilon_{x\pm}=\frac{z^2}{2\mu} \partial_x p_{\pm} + a_{x\pm} z + b_{x\pm}, 
\quad 
\upsilon_{y\pm}=\frac{z^2}{2\mu} \partial_yp_\pm + a_{y\pm} z + b_{y\pm},
\end{equation}
where $\pm$ denotes the liquid above or below the membrane.
The ten quantities $p_\pm,a_{x\pm},a_{y\pm},b_{x\pm},b_{y\pm}$
are functions of $x$ and $y$, and are constants with respect to $z$.
The fluid velocity at the two walls and at the membrane satisfy a no-slip condition
\begin{align}
\label{aeq:bc1}
\upsilon_{x\pm}\rvert_{z=\pm h_0}=\pm \upsilon_0, \quad \upsilon_{y\pm}\rvert_{z=\pm h_0}=0,
\\
\label{aeq:bc2}
\upsilon_{x+}\rvert_{z= h}=\upsilon_{x-}\rvert_{z= h}, 
\quad 
\upsilon_{y+}\rvert_{z= h}=\upsilon_{y-}\rvert_{z= h}.
\end{align}
Moreover, the tangential stresses are assumed to be
continuous across the membrane 
\begin{align}
\label{aeq:bc3}
\partial_z\upsilon_{x+}\rvert_{z=h}=\partial_z\upsilon_{x-}\rvert_{z=h}, \quad 
\partial_z\upsilon_{y+}\rvert_{z=h}=\partial_z\upsilon_{y-}\rvert_{z=h}.
\end{align}
In addition, the normal force $f_z$ which is oriented along $z$ to leading order 
balances the difference of pressure above and below the membrane
\begin{align}
\label{aeq:bc4}
p_+-p_- = f_z.
\end{align}

\subsection{Tangential forces}
The tangential forces on the walls are due to the viscous 
shear stress exerted by the fluid on the wall.
The two components of the friction force per unit area 
are given by the difference of shear stresses on the upper
 and lower walls
\begin{align}
\label{eq:viscousforce}
&f_{wx}=\frac{\mu}{{\cal L}_x{\cal L}_y} \int_0^{{\cal L}_x} dx \int_0^{{\cal L}_y} dy 
\left( \partial_z \upsilon_{x+}\rvert_{z=+ h_0}+\partial_z \upsilon_{x-}\rvert_{z=- h_0} \right), \nonumber \\
&f_{wy}=\frac{\mu}{{\cal L}_x{\cal L}_y} \int_0^{{\cal L}_x} dx \int_0^{{\cal L}_y} dy 
\left( \partial_z \upsilon_{y+}\rvert_{z=+ h_0}+\partial_z \upsilon_{y-}\rvert_{z=- h_0} \right).
\end{align}
Substituting the expression of the hydrodynamic velocity Eq. (\ref{aeq:poiseuille})
in the boundary conditions Eqs. (\ref{aeq:bc1}-\ref{aeq:bc4})
and combining these expressions, we obtain 
\begin{align}
\left( \partial_z \upsilon_{x+}\rvert_{z=+ h_0}+\partial_z \upsilon_{x-}\rvert_{z=- h_0} \right)
=2\frac{v_0}{h_0}+\frac{h^2-h_0^2}{2\mu h_0}\partial_xf_z, \nonumber \\
\left( \partial_z \upsilon_{y+}\rvert_{z=+ h_0}+\partial_z \upsilon_{y-}\rvert_{z=- h_0} \right)
=\frac{h^2-h_0^2}{2\mu h_0}\partial_x f_z.
\end{align}
Inserting these expressions into Eq. (\ref{eq:viscousforce}) leads to 
Eqs. (\ref{eq:totalforce2D},\ref{eq:membrane_force_2D}).

\subsection{Dynamical equation}
In order to determine the ten functions $p_\pm,a_{x\pm},a_{y\pm},b_{x\pm},b_{y\pm}$,
the seven boundary conditions Eqs. (\ref{aeq:bc1}-\ref{aeq:bc4}) are not sufficient.

Additional boundary conditions at the walls and at the
membrane involve vertical flow.
The vertical motion of the interface is associated to 
vertical hydrodynamic flows
\begin{align}
\label{aeq:kinematic_continuity_z}
    u_{z+}\rvert_{z=h}=\partial_th, 
    \quad
    u_{z-}\rvert_{z=h}=\partial_th.
\end{align}
Vertical hydrodynamic flows at the walls 
are caused by flow through the porous walls~~\cite{membraneadhesion2D2018,LeGoff2014}
\begin{align}
\label{aeq:BC_permeability}
    u_{z+}\rvert_{z=+h_0}= \nu (p_+ - p_{eq}),
    \quad
    u_{z-}\rvert_{z=-h_0}= -\nu (p_- - p_{eq}),
\end{align}
where $p_{eq}$ is a reference pressure.

In the lubrication expansion, vertical flow in the $z$ direction is smaller
than the flow in the $x,y$ plane, and appears to higher order. 
However, vertical flows are crucial and balance the large-scale
variations of the horizontal flow in mass conservation.

Since we have four additional relations Eqs. (\ref{aeq:kinematic_continuity_z},\ref{aeq:BC_permeability}),
we have in total eleven relations and ten unknowns. We therefore
find the expression of all the unknown, plus one evolution relating $\partial_th$
and the other physical quantities.
This leads to complex expressions in general, and detailed
derivations are reported in Ref.~\cite{membraneadhesion2D2018,LeGoff2014}.
In the limit of large $\bar\nu$, one obtains Eq. (\ref{eq:dynamic2D}).

\subsection{Limit $\bar\nu\gg 1$}

Here, we report a heuristic discussion of the 
meaning of the parameter $\bar\nu$, and of the limit $\bar\nu\gg 1$.
Let us discard the shear flow created by the motion of the walls for simplicity.
First, we notice that since the hydrodynamic velocity in the $(x,y)$
plane is a Poiseuille flow, which is quadratic in $z$ from Eq. (\ref{aeq:poiseuille}),
the total flow $\mathbf j=(j_x,j_y)$ integrated over $z$ is cubic in $h_0$, i.e.
$j\sim (h_0^3/\mu)\nabla_{xy}p$.
The motion of the membrane can either result 
from a non-constant flow above or below, leading to $\partial_th\sim -\nabla_{xy}\cdot \mathbf j$,
or from the loss of mass through the permeable walls, leading to $\partial_th\sim\nu(p-p_{eq})$.
These correspond to two different dissipation mechanisms controlled by the 
kinetic coefficients $\mu$ and $\nu$.
Assume now that we consider a membrane pattern with a lengthscale $\ell$
in the $x,y$ plane. Then, $\nabla_{xy}\sim 1/\ell$, and $\nabla_{xy}\cdot \mathbf j\sim (h_0^3/\mu)p/\ell^2$.
Hence, the motion of the membrane is limited by hydrodynamic flow parallel
to the walls when $(h_0^3/\mu)p/\ell^2\gg\nu(p-p_{eq})$, and limited by
the flow through the walls in the opposite limit. Eliminating $p$,
we obtain viscosity-limited motion for $\ell\ll\ell_{kin}$
and permeability-limited regime for $\ell\gg\ell_{kin}$, where
\begin{align}
\ell_{kin}=\left(\frac{h_0^3}{\mu \nu}\right)^{1/2} .
\end{align}

Choosing the natural lengthscale of the pattern $\ell\sim\ell_\parallel$ defined in Eq. (\ref{eq:normalisationdistance}),
we obtain the viscosity-limited motion for $\bar\nu\ll 1$ 
and permeability-limited regime for $\bar\nu\gg 1$,
where $\bar\nu$ is defined in Eq. (\ref{eq:barnu_def}).

Two remarks are in order. First, our analysis in this paper focuses
on the permeability-limited regime for $\bar\nu\gg 1$ which corresponds
to patterns that exhibit a lengthscale larger than $\ell_{kin}$.
Second, the parameter $\nu$ does not compare vertical and horizontal hydrodynamic velocities.
Indeed, from  Eq. (\ref{aeq:poiseuille}), horizontal velocities scale as $h_0^2p/(\mu\ell)$,
vertical velocities scale as $\nu(p-p{eq})$. Their ratio reads 
$\ell\mu\nu/h_0^2\sim \bar\nu (h_0/\ell)$. Since $\epsilon=h_0/\ell$
is small from the very definition of the lubrication approximation, 
the vertical hydrodynamic velocities are always smaller than the horizontal ones.
In addition, the strict validity of the lubrication approximation requires
that $\bar\nu\epsilon$ should be small even when $\bar\nu$ is large, i.e.
that $\epsilon$ is smaller than $\bar\nu^{-1}$.

\section{Sine-profile ansatz without shear}
\label{secapp:sineansatz_noshear}

\subsection{General procedure}

Let $\zeta$  be the variable in the direction perpendicular to the wrinkles. 
We assume that the membrane profile is sinusoidal
$H(\zeta)=a\cos(q\zeta)$ where $a$ is the amplitude and $q$ is the wave number $q=2\pi/\lambda$, 
$\lambda$ is the wavelength of the wrinkle.
Using this ansatz the relation between the amplitude $a$ and 
the root-mean-square amplitude $\langle H^2 \rangle$ is $a=\sqrt{2}\langle H^2 \rangle^\frac{1}{2}$.

Consider the free energy $\xi$ per unit length of one straight wrinkle:
\begin{equation}
\xi=\int_0^\lambda d\zeta \left\{ \frac{1}{2}(\partial_{\zeta \zeta}H)^2 +\frac{\Sigma_0}{2}(\partial_\zeta H)^2 + U(H) \right\},
\end{equation}
where the tension $\Sigma_0$ accounts for the constraint of fixed total excess area.
Substituting the sine ansatz, we obtain
\begin{equation}
\xi=\frac{2\pi}{q} \left\{ \frac{1}{4}(a^2q^4+a^2q^2\Sigma_0) + J_U \right\}.
\end{equation}
The free energy density (i.e. the free energy per unit area) is then given by
\begin{equation}
\bar{\xi}=\frac{\xi}{\lambda}=\frac{1}{4}(a^2q^4+a^2q^2\Sigma_0) + J_U,
\end{equation}
where
\begin{equation}
J_U=\frac{1}{2\pi} \int_0^{2\pi} d\eta U(H).
\end{equation}
Here we have defined the integration variable $\eta=q\zeta$.

To find the wavelength $\tilde{\lambda}=2\pi/\tilde{q}$ and the amplitude $\tilde{a}$ 
that minimize the energy density $\bar{\xi}$, we first minimize $\bar{\xi}$ with respect to $q$. 
Solving $\partial_q\bar{\xi}=0$ we obtain the relation $\Sigma_0=-2\tilde{q}^2$. 
 Thus, we have
\begin{equation}
\left. \bar{\xi} \right \rvert _{q=\tilde{q}} = -\frac{1}{4} a^2 \tilde{q}^4 + J_U.
\end{equation}
Next we minimize $\left. \bar{\xi} \right \rvert _{q=\tilde{q}}$ with respect to $a$. 
Solving $\partial_a \left( \left. \bar{\xi} \right \rvert _{q=\tilde{q}}\right)=0$ 
we find the dependence of $\tilde{q}$ on $\tilde{a}$
\begin{equation}
\label{eq:wavenumber}
\tilde{q}=\left[ \frac{2}{\tilde{a}} \left. ( \partial_a J_U ) \right \rvert _{a=\tilde{a}} \right]^\frac{1}{4}.
\end{equation}
To establish this relation, we have used the fact that since $U(H)$ depends only on $H$ and not on its spatial derivatives, 
$J_U$ depends only on $a$, and not on $q$.
Note in addition that we could have also obtained this relation
by direct substitution of the relation $\Sigma_0=-2\tilde{q}^2$ into Eq. (\ref{eq:sigma0_normalized}).

A relation between $\tilde{a}$ and the membrane excess area $\Delta A_*$ is then obtained from 
evaluation of the excess area with the sine-profile ansatz, 
\begin{equation}
\label{eq:Delta_A*_vs_a}
\Delta A_*=\left(\frac{\tilde{q}\tilde{a}}{2}\right)^2
=\left[ \frac{\tilde{a}^3}{8} \left. ( \partial_a J_U ) \right \rvert _{a=\tilde{a}} \right]^\frac{1}{2}.
\end{equation}
As a summary, the amplitude $a$ can be calculated
from the  excess area $\Delta A_*$ from the inversion of Eq. (\ref{eq:Delta_A*_vs_a}).
Then, the wavelength of the wrinkles is obtained as a function of $a$
using Eq. (\ref{eq:wavenumber}).

\subsection{Case of the potential $U(H)=(1-H^p)^{-m}$}

\begin{figure}
\begin{center}
    {\includegraphics[width=0.9\textwidth]{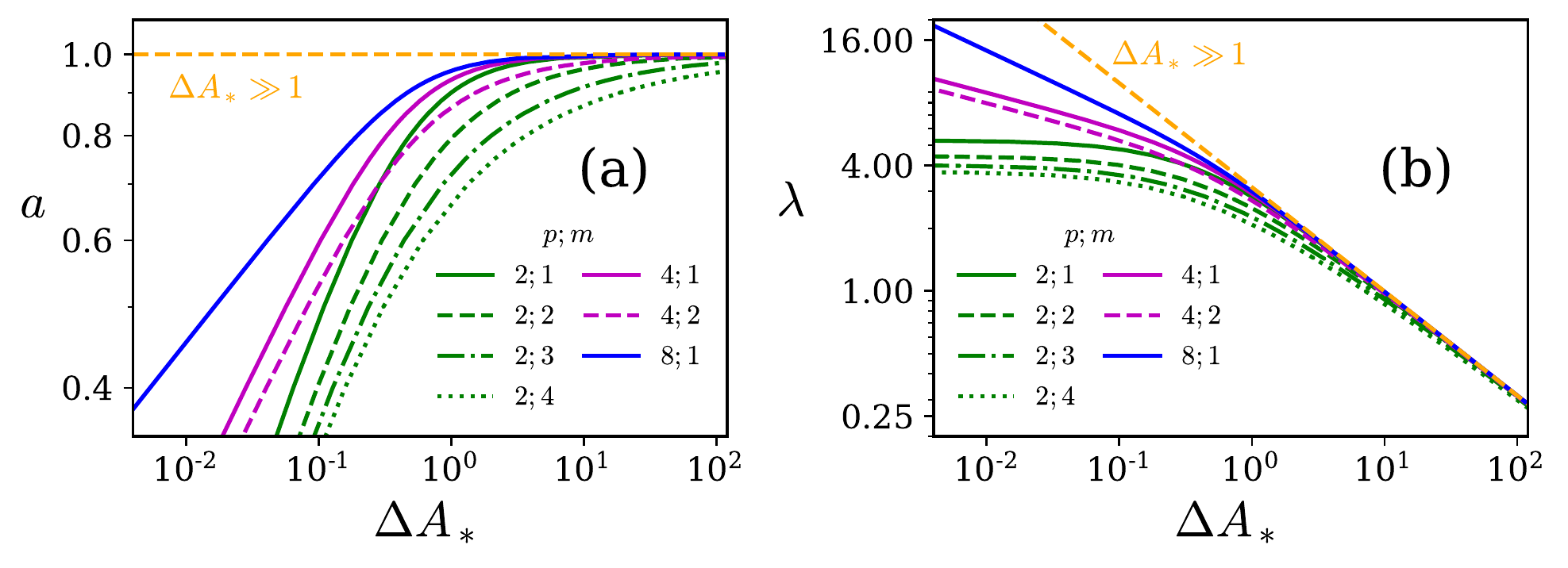}}
\caption{
{
Predictions of the sine-ansatz Eq. (\ref{eq:qgeneral}).
(a) Wrinkle amplitude $a$ and (b) wrinkle wavelength $\lambda$} of the membrane profile as a function of the normalized excess area for various values of $m$ and $p$
using the potential $U(H)=(1-H^p)^{-m}$. 
The yellow dashed lines show the analytical results for large excess area Eq. (\ref{eq:lambda_Alarge}): $a=1$ and $\lambda=\pi \Delta A_*^{-1/2}$.
}
\label{fig:avsA_mp}
\end{center}
\end{figure}

In order to find $\tilde{q}$ and $\tilde{a}$ more explicitly we need to calculate $J_U$ 
and $\partial_a J_U$ for a given $U(H)$. Consider the energy potential of the general form
\begin{equation}
U(H)=(1-H^p)^{-m},
\end{equation}
where $p$ is a positive even integer and $m$ is a positive integer. 
We then have
\begin{align}
J_U &=1+\sum_{k=1}^\infty {m+k-1 \choose k} I_{pk}a^{pk},
\nonumber \\ 
\partial_a J_U &=\frac{p}{a}\sum_{k=1}^\infty {m+k-1 \choose k} kI_{pk}a^{pk}.
\end{align}
We have defined
\begin{equation}
I_{pk}=\frac{1}{2\pi} \int_0^{2\pi} d\eta \cos^{pk}\eta=\frac{1}{\pi}B\left(\frac{1}{2},\frac{1}{2}+\frac{pk}{2} \right),
\end{equation}
where $B(x,y)$ is the Beta function. 
Finally, we obtain
\begin{align}
\label{eq:qgeneral}
\tilde{q}&=\left[ \frac{2p}{\tilde{a}^2}\sum_{k=1}^\infty {m+k-1 \choose k} kI_{pk}\tilde{a}^{pk} \right]^\frac{1}{4}, \nonumber \\
\Delta A_*&=\left[ \frac{p}{8}\sum_{k=1}^\infty {m+k-1 \choose k} kI_{pk}\tilde{a}^{pk+2} \right]^\frac{1}{2}.
\end{align}
The results of Eqs. (\ref{eq:qgeneral}) is shown in Fig. \ref{fig:avsA_mp} for some values of $m$ and $p$.

\subsection{Limit of small amplitudes for the potential $U(H)=(1-H^p)^{-m}$}
For small amplitudes $a$, we only keep the first term in the series expansion of Eq. (\ref{eq:qgeneral}) 
\begin{align}
\tilde{q}&=\left(2mpI_p\tilde{a}^{p-2} \right)^\frac{1}{4},
\nonumber \\
\Delta A_*&=\left( \frac{pm}{8}  I_{p}\tilde{a}^{p+2} \right)^\frac{1}{2},
\end{align}
where we have defined the constant
\begin{equation}
I_p=I_{pk}\rvert_{k=1}=\frac{1}{\pi}B\left(\frac{1}{2},\frac{1}{2}+\frac{p}{2} \right).
\end{equation}
We finally  obtain
\begin{align}
\label{eq:sineansatzsmalla}
\tilde{a}&=\Delta A_*^\frac{2}{p+2} C_{m,p},\nonumber \\
\tilde{q}&=2\Delta A_*^{\frac{1}{2}-\frac{2}{p+2}} (C_{m,p})^{-1},\nonumber \\
\tilde{\lambda}&=\frac{2\pi}{\tilde{q}}= \Delta A_* ^{-\frac{1}{2}+\frac{2}{p+2}} \pi C_{m,p},
\end{align}
where we have defined
\begin{equation}
\label{eq:Cmp}
C_{m,p}=\left( \frac{8}{mpI_p} \right)^\frac{1}{p+2}.
\end{equation}

\section{Random initial conditions}
\label{a:random_init_cond}

Our simulation scheme with area conservation requires a random smooth initial condition for the excess area to be well defined. Such an initial condition with excess area $\Delta A_*$ can be obtained by solving the time-dependent Ginzburg-Landau (TDGL) equation
\begin{equation}
\partial_T H=w^2\Delta H-U_\mathrm{adh}^\prime(H),
\end{equation}
using an explicit scheme with finite-differences and random initial conditions. 
We use a double-well adhesion potential~\cite{membraneadhesion2D2018}
\begin{equation}
U_\mathrm{adh}=\frac{1}{4}(H_m^2-H^2)^2.
\end{equation}
Membrane profiles with different excess areas $\Delta A_*$ can be selected (as initial conditions for the main simulations) by varying the potential well $H_m$ ($H_m<1$), the domain width $w$ (typically $w<1$), and the simulation time of the TDGL equation (typically between 10 and 30).
In our previous work, we have obtained evidences that 
such a procedure does not affect the details of the subsequent dynamics \cite{membraneadhesion2D2018}. 

\section{Numerical evaluation of the drift velocity}
\label{secapp:Vdriftapp}

In this appendix, we report on the numerical evaluation
of the drift velocity of drifting steady-states
$\mathbf{V}_d$ with its two components $V_{dX}$ and $V_{dY}$.
In order to validate the method, we first perform a
direct measurement on the images of the drift velocity
of some defect. We refer to this method as Method 1 and denote the velocity as $V_{d1X}$ and $V_{d1Y}$ of the membrane profile.

The drift velocity can also be evaluated numerically as follows. 
A drifting steady-state obeys
\begin{equation}
\label{eq:driftsteady}
-V_{dX}\partial_XH -V_{dY}\partial_YH=F_Z - VH\partial_XH.
\end{equation}
We multiply Eq. (\ref{eq:driftsteady}) with an arbitrary function $\phi(H)$ and integrate over $X$ to obtain
\begin{equation}
\label{eq:vdy}
V_{dY} =-\frac{\int dX [\phi(H)F_Z] }{ \int dX [\phi(H)\partial_YH]}.
\end{equation}
Likewise, to calculate $V_{dX}$ we multiply Eq. (\ref{eq:driftsteady}) with an arbitrary function $\chi(H)$ and integrate over $Y$ to get
\begin{equation}
\label{eq:vdx}
V_{dX} =-\frac{\int dY [\chi(H)F_Z] + V\int dY [\chi(H)H\partial_XH]}{\int dY [\chi(H)\partial_XH]}.
\end{equation}
Note that Eq. (\ref{eq:vdy}) provides an evaluation of $V_{dY}$ for
each value of $X$, and Eq. (\ref{eq:vdx}) provides an evaluation of $V_{dX}$ for
each value of $Y$. The main difficulty of this method is that the denominator
of Eqs. (\ref{eq:vdy},\ref{eq:vdx}) is very small for some specific values of $Y$ or $X$,
leading to numerical inaccuracies or divergences. A simple solution
to this difficulty is to evaluate these expressions for all $Y$ and $X$ 
on our numerical spatial grid,
and take the median value.
We find that the resulting estimate is reliable,
as shown by the comparison with direct measurements in Table 1.

\begin{table*}
\centering
\begin{tabular}{|c|c|c|c|c|c|c|}
\hline
$\Delta A_*$ & $V$ & $T$ & $V_{d2X}$ & $V_{d1X}$ & $V_{d2Y}$ & $V_{d1Y}$ \\ \hline
$0.007495$ & 0.01 & $1.99\times 10^6$ & $6.8545\times 10^{-6}$ & $7.07\times 10^{-6}$ & $1.5721\times 10^{-6}$ & $2.02\times 10^{-6}$ \\ \hline
$0.007495$ & 0.1 & $1.45\times 10^6$ & $5.7022\times 10^{-5}$ & $5.634\times 10^{-5}$ & $1.7475\times 10^{-5}$ & $1.878\times 10^{-5}$ \\ \hline
$0.007495$ & 0.5 & $5\times 10^5$ & $-1.3988\times 10^{-3}$ & $-1.4\times 10^{-3}$ & $-5.0391\times 10^{-5}$ & $-5.13\times 10^{-5}$ \\ \hline
$0.007495$ & 1 & $2\times 10^5$ & $7.3187\times 10^{-3}$ & $7.36\times 10^{-3}$ & $-1.8964\times 10^{-4}$ & $-2.32\times 10^{-4}$  \\ \hline
$0.01186$ & 0.1 & $5\times 10^5$ & $1.0634\times 10^{-5}$ & $1.05\times 10^{-5}$ & $-6.1509\times 10^{-6}$ & $-6.31\times 10^{-6}$ \\ \hline
$0.01186$ & 0.2 & $1.28\times 10^6$ & $5.5289\times10^{-5}$ & $6.25\times 10^{-5}$ & $1.8618\times 10^{-5}$ & $0.947\times 10^{-5}$ \\ \hline
$0.1503$ & 0.5 & $5\times 10^5$ & $3.9351\times 10^{-5}$ & $3.33\times 10^{-5}$ & $4.5404\times 10^{-7}$ & $0$ \\ \hline
$0.1503$ & 1 & $4.8\times 10^5$ & $4.853\times 10^{-4}$ & $4.7083\times 10^{-4}$ & $4.2342\times 10^{-5}$ & $4.167\times 10^{-5}$ \\ \hline
$0.1503$ & 2 & $5\times 10^5$ & $6.6394\times 10^{-4}$ & $6.18\times 10^{-4}$ & $-8.6122\times 10^{-5}$ & $-8.94\times 10^{-5}$ \\ \hline
$0.1503$ & 10 & $5\times 10^5$ & $8.0422\times 10^{-4}$ & $8.42\times 10^{-4}$ & $-4.8108\times 10^{-7}$ & $0$ \\ \hline
$0.2462$ & 5 & $4\times 10^5$ & $-5.0496\times 10^{-4}$ & $-6.24\times 10^{-4}$ & $-2.2137\times 10^{-5}$ & $-1.44\times 10^{-5}$ \\ \hline
$0.2462$ & 20 & $1.5\times 10^5$ & $2.2826\times 10^{-3}$ & $2.2279\times 10^{-3}$ & $6.196\times 10^{-7}$ & $0$ \\ \hline
\end{tabular}
\caption{Drift velocity along $X$ and $Y$. 
$V_{d1X}$ and $V_{d1Y}$ are values from manual measurement from images. $V_{d2X}$ and $V_{d2Y}$ are medians of the calculated values using $\phi(H)=\chi(H)=1$.}
\end{table*}

\end{appendix}

\bibliography{membraneshear2D_ref}

\begin{thebibliography}{38}%
\makeatletter
\providecommand \@ifxundefined [1]{%
 \@ifx{#1\undefined}
}%
\providecommand \@ifnum [1]{%
 \ifnum #1\expandafter \@firstoftwo
 \else \expandafter \@secondoftwo
 \fi
}%
\providecommand \@ifx [1]{%
 \ifx #1\expandafter \@firstoftwo
 \else \expandafter \@secondoftwo
 \fi
}%
\providecommand \natexlab [1]{#1}%
\providecommand \enquote  [1]{``#1''}%
\providecommand \bibnamefont  [1]{#1}%
\providecommand \bibfnamefont [1]{#1}%
\providecommand \citenamefont [1]{#1}%
\providecommand \href@noop [0]{\@secondoftwo}%
\providecommand \href [0]{\begingroup \@sanitize@url \@href}%
\providecommand \@href[1]{\@@startlink{#1}\@@href}%
\providecommand \@@href[1]{\endgroup#1\@@endlink}%
\providecommand \@sanitize@url [0]{\catcode `\\12\catcode `\$12\catcode
  `\&12\catcode `\#12\catcode `\^12\catcode `\_12\catcode `\%12\relax}%
\providecommand \@@startlink[1]{}%
\providecommand \@@endlink[0]{}%
\providecommand \url  [0]{\begingroup\@sanitize@url \@url }%
\providecommand \@url [1]{\endgroup\@href {#1}{\urlprefix }}%
\providecommand \urlprefix  [0]{URL }%
\providecommand \Eprint [0]{\href }%
\providecommand \doibase [0]{http://dx.doi.org/}%
\providecommand \selectlanguage [0]{\@gobble}%
\providecommand \bibinfo  [0]{\@secondoftwo}%
\providecommand \bibfield  [0]{\@secondoftwo}%
\providecommand \translation [1]{[#1]}%
\providecommand \BibitemOpen [0]{}%
\providecommand \bibitemStop [0]{}%
\providecommand \bibitemNoStop [0]{.\EOS\space}%
\providecommand \EOS [0]{\spacefactor3000\relax}%
\providecommand \BibitemShut  [1]{\csname bibitem#1\endcsname}%
\let\auto@bib@innerbib\@empty
\bibitem [{\citenamefont {Trunfio-Sfarghiu}\ \emph {et~al.}(2008)\citenamefont
  {Trunfio-Sfarghiu}, \citenamefont {Berthier}, \citenamefont {Meurisse},\ and\
  \citenamefont {Rieu}}]{JeanPaulRieu2008}%
  \BibitemOpen
  \bibfield  {author} {\bibinfo {author} {\bibfnamefont {A.-M.}\ \bibnamefont
  {Trunfio-Sfarghiu}}, \bibinfo {author} {\bibfnamefont {Y.}~\bibnamefont
  {Berthier}}, \bibinfo {author} {\bibfnamefont {M.-H.}\ \bibnamefont
  {Meurisse}}, \ and\ \bibinfo {author} {\bibfnamefont {J.-P.}\ \bibnamefont
  {Rieu}},\ }\href {\doibase 10.1021/la8005234} {\bibfield  {journal} {\bibinfo
   {journal} {Langmuir}\ }\textbf {\bibinfo {volume} {24}},\ \bibinfo {pages}
  {8765} (\bibinfo {year} {2008})}\BibitemShut {NoStop}%
\bibitem [{\citenamefont {Schmidt}\ \emph {et~al.}(2007)\citenamefont
  {Schmidt}, \citenamefont {Gastelum}, \citenamefont {Nguyen}, \citenamefont
  {Schumacher},\ and\ \citenamefont {Sah}}]{Schmidt2007}%
  \BibitemOpen
  \bibfield  {author} {\bibinfo {author} {\bibfnamefont {T.~A.}\ \bibnamefont
  {Schmidt}}, \bibinfo {author} {\bibfnamefont {N.~S.}\ \bibnamefont
  {Gastelum}}, \bibinfo {author} {\bibfnamefont {Q.~T.}\ \bibnamefont
  {Nguyen}}, \bibinfo {author} {\bibfnamefont {B.~L.}\ \bibnamefont
  {Schumacher}}, \ and\ \bibinfo {author} {\bibfnamefont {R.~L.}\ \bibnamefont
  {Sah}},\ }\href {\doibase 10.1002/art.22446} {\bibfield  {journal} {\bibinfo
  {journal} {Arthritis and Rheumatology}\ }\textbf {\bibinfo {volume} {56}},\
  \bibinfo {pages} {882} (\bibinfo {year} {2007})}\BibitemShut {NoStop}%
\bibitem [{\citenamefont {Botan}\ \emph {et~al.}(2015)\citenamefont {Botan},
  \citenamefont {Joly}, \citenamefont {Fillot},\ and\ \citenamefont
  {Loison}}]{Loison2015}%
  \BibitemOpen
  \bibfield  {author} {\bibinfo {author} {\bibfnamefont {A.}~\bibnamefont
  {Botan}}, \bibinfo {author} {\bibfnamefont {L.}~\bibnamefont {Joly}},
  \bibinfo {author} {\bibfnamefont {N.}~\bibnamefont {Fillot}}, \ and\ \bibinfo
  {author} {\bibfnamefont {C.}~\bibnamefont {Loison}},\ }\href {\doibase
  10.1021/acs.langmuir.5b02786} {\bibfield  {journal} {\bibinfo  {journal}
  {Langmuir}\ }\textbf {\bibinfo {volume} {31}},\ \bibinfo {pages} {12197}
  (\bibinfo {year} {2015})}\BibitemShut {NoStop}%
\bibitem [{\citenamefont {Swann}\ \emph {et~al.}(1984)\citenamefont {Swann},
  \citenamefont {Bloch}, \citenamefont {Swindell},\ and\ \citenamefont
  {Shore}}]{Swann1984}%
  \BibitemOpen
  \bibfield  {author} {\bibinfo {author} {\bibfnamefont {D.~A.}\ \bibnamefont
  {Swann}}, \bibinfo {author} {\bibfnamefont {K.~J.}\ \bibnamefont {Bloch}},
  \bibinfo {author} {\bibfnamefont {D.}~\bibnamefont {Swindell}}, \ and\
  \bibinfo {author} {\bibfnamefont {E.}~\bibnamefont {Shore}},\ }\href
  {\doibase 10.1002/art.1780270511} {\bibfield  {journal} {\bibinfo  {journal}
  {Arthritis and Rheumatology}\ }\textbf {\bibinfo {volume} {27}},\ \bibinfo
  {pages} {552} (\bibinfo {year} {1984})}\BibitemShut {NoStop}%
\bibitem [{\citenamefont {Tadmor}\ \emph {et~al.}(2002)\citenamefont {Tadmor},
  \citenamefont {Chen},\ and\ \citenamefont {Israelachvili}}]{Tadmor2002}%
  \BibitemOpen
  \bibfield  {author} {\bibinfo {author} {\bibfnamefont {R.}~\bibnamefont
  {Tadmor}}, \bibinfo {author} {\bibfnamefont {N.}~\bibnamefont {Chen}}, \ and\
  \bibinfo {author} {\bibfnamefont {J.~N.}\ \bibnamefont {Israelachvili}},\
  }\href {\doibase 10.1002/jbm.10215} {\bibfield  {journal} {\bibinfo
  {journal} {J. Biomedical Materials Research}\ }\textbf {\bibinfo {volume}
  {61}},\ \bibinfo {pages} {514} (\bibinfo {year} {2002})}\BibitemShut
  {NoStop}%
\bibitem [{\citenamefont {Erdemir}(2005)}]{Erdemir2005}%
  \BibitemOpen
  \bibfield  {author} {\bibinfo {author} {\bibfnamefont {A.}~\bibnamefont
  {Erdemir}},\ }\href {\doibase https://doi.org/10.1016/j.triboint.2004.08.008}
  {\bibfield  {journal} {\bibinfo  {journal} {Tribology International}\
  }\textbf {\bibinfo {volume} {38}},\ \bibinfo {pages} {249 } (\bibinfo {year}
  {2005})},\ \bibinfo {note} {boundary Lubrication}\BibitemShut {NoStop}%
\bibitem [{\citenamefont {Raviv}\ \emph {et~al.}(2003)\citenamefont {Raviv},
  \citenamefont {Giasson}, \citenamefont {Kampf}, \citenamefont {Gohy},
  \citenamefont {J{\'e}r{\^o}me},\ and\ \citenamefont {Klein}}]{Raviv2003}%
  \BibitemOpen
  \bibfield  {author} {\bibinfo {author} {\bibfnamefont {U.}~\bibnamefont
  {Raviv}}, \bibinfo {author} {\bibfnamefont {S.}~\bibnamefont {Giasson}},
  \bibinfo {author} {\bibfnamefont {N.}~\bibnamefont {Kampf}}, \bibinfo
  {author} {\bibfnamefont {J.-F.}\ \bibnamefont {Gohy}}, \bibinfo {author}
  {\bibfnamefont {R.}~\bibnamefont {J{\'e}r{\^o}me}}, \ and\ \bibinfo {author}
  {\bibfnamefont {J.}~\bibnamefont {Klein}},\ }\href {\doibase
  10.1038/nature01970} {\bibfield  {journal} {\bibinfo  {journal} {Nature}\
  }\textbf {\bibinfo {volume} {425}},\ \bibinfo {pages} {163} (\bibinfo {year}
  {2003})}\BibitemShut {NoStop}%
\bibitem [{\citenamefont {Briscoe}\ \emph {et~al.}(2006)\citenamefont
  {Briscoe}, \citenamefont {Titmuss}, \citenamefont {Tiberg}, \citenamefont
  {Thomas}, \citenamefont {McGillivray},\ and\ \citenamefont
  {Klein}}]{Briscoe2006}%
  \BibitemOpen
  \bibfield  {author} {\bibinfo {author} {\bibfnamefont {W.~H.}\ \bibnamefont
  {Briscoe}}, \bibinfo {author} {\bibfnamefont {S.}~\bibnamefont {Titmuss}},
  \bibinfo {author} {\bibfnamefont {F.}~\bibnamefont {Tiberg}}, \bibinfo
  {author} {\bibfnamefont {R.~K.}\ \bibnamefont {Thomas}}, \bibinfo {author}
  {\bibfnamefont {D.~J.}\ \bibnamefont {McGillivray}}, \ and\ \bibinfo {author}
  {\bibfnamefont {J.}~\bibnamefont {Klein}},\ }\href {\doibase
  10.1038/nature05196} {\bibfield  {journal} {\bibinfo  {journal} {Nature}\
  }\textbf {\bibinfo {volume} {444}},\ \bibinfo {pages} {191} (\bibinfo {year}
  {2006})}\BibitemShut {NoStop}%
\bibitem [{\citenamefont {Gov}\ \emph {et~al.}(2004)\citenamefont {Gov},
  \citenamefont {Zilman},\ and\ \citenamefont {Safran}}]{Gov2004}%
  \BibitemOpen
  \bibfield  {author} {\bibinfo {author} {\bibfnamefont {N.}~\bibnamefont
  {Gov}}, \bibinfo {author} {\bibfnamefont {A.~G.}\ \bibnamefont {Zilman}}, \
  and\ \bibinfo {author} {\bibfnamefont {S.}~\bibnamefont {Safran}},\ }\href
  {\doibase 10.1103/PhysRevE.70.011104} {\bibfield  {journal} {\bibinfo
  {journal} {Phys. Rev. E}\ }\textbf {\bibinfo {volume} {70}},\ \bibinfo
  {pages} {011104} (\bibinfo {year} {2004})}\BibitemShut {NoStop}%
\bibitem [{\citenamefont {Marlow}\ and\ \citenamefont
  {Olmsted}(2002)}]{Olmstead2002}%
  \BibitemOpen
  \bibfield  {author} {\bibinfo {author} {\bibfnamefont {S.~W.}\ \bibnamefont
  {Marlow}}\ and\ \bibinfo {author} {\bibfnamefont {P.~D.}\ \bibnamefont
  {Olmsted}},\ }\href {\doibase 10.1103/PhysRevE.66.061706} {\bibfield
  {journal} {\bibinfo  {journal} {Phys. Rev. E}\ }\textbf {\bibinfo {volume}
  {66}},\ \bibinfo {pages} {061706} (\bibinfo {year} {2002})}\BibitemShut
  {NoStop}%
\bibitem [{\citenamefont {Huang}\ and\ \citenamefont {Im}(2006)}]{Huang2006}%
  \BibitemOpen
  \bibfield  {author} {\bibinfo {author} {\bibfnamefont {R.}~\bibnamefont
  {Huang}}\ and\ \bibinfo {author} {\bibfnamefont {S.~H.}\ \bibnamefont {Im}},\
  }\href {\doibase 10.1103/PhysRevE.74.026214} {\bibfield  {journal} {\bibinfo
  {journal} {Phys. Rev. E}\ }\textbf {\bibinfo {volume} {74}},\ \bibinfo
  {pages} {026214} (\bibinfo {year} {2006})}\BibitemShut {NoStop}%
\bibitem [{\citenamefont {To}\ \emph {et~al.}(2018)\citenamefont {To},
  \citenamefont {Le~Goff},\ and\ \citenamefont
  {Pierre-Louis}}]{membraneadhesion2D2018}%
  \BibitemOpen
  \bibfield  {author} {\bibinfo {author} {\bibfnamefont {T.~B.~T.}\
  \bibnamefont {To}}, \bibinfo {author} {\bibfnamefont {T.}~\bibnamefont
  {Le~Goff}}, \ and\ \bibinfo {author} {\bibfnamefont {O.}~\bibnamefont
  {Pierre-Louis}},\ }\href {\doibase 10.1039/C8SM01567H} {\bibfield  {journal}
  {\bibinfo  {journal} {Soft Matter}\ }\textbf {\bibinfo {volume} {14}},\
  \bibinfo {pages} {8552} (\bibinfo {year} {2018})}\BibitemShut {NoStop}%
\bibitem [{\citenamefont {Cerda}\ and\ \citenamefont
  {Mahadevan}(2003)}]{Cerda2003}%
  \BibitemOpen
  \bibfield  {author} {\bibinfo {author} {\bibfnamefont {E.}~\bibnamefont
  {Cerda}}\ and\ \bibinfo {author} {\bibfnamefont {L.}~\bibnamefont
  {Mahadevan}},\ }\href {\doibase 10.1103/PhysRevLett.90.074302} {\bibfield
  {journal} {\bibinfo  {journal} {Phys. Rev. Lett.}\ }\textbf {\bibinfo
  {volume} {90}},\ \bibinfo {pages} {074302} (\bibinfo {year}
  {2003})}\BibitemShut {NoStop}%
\bibitem [{\citenamefont {Leocmach}\ \emph {et~al.}(2015)\citenamefont
  {Leocmach}, \citenamefont {Nespoulous}, \citenamefont {Manneville},\ and\
  \citenamefont {Gibaud}}]{LeocmachSciAdv2015}%
  \BibitemOpen
  \bibfield  {author} {\bibinfo {author} {\bibfnamefont {M.}~\bibnamefont
  {Leocmach}}, \bibinfo {author} {\bibfnamefont {M.}~\bibnamefont
  {Nespoulous}}, \bibinfo {author} {\bibfnamefont {S.}~\bibnamefont
  {Manneville}}, \ and\ \bibinfo {author} {\bibfnamefont {T.}~\bibnamefont
  {Gibaud}},\ }\href {\doibase 10.1126/sciadv.1500608} {\bibfield  {journal}
  {\bibinfo  {journal} {Science Advances}\ }\textbf {\bibinfo {volume} {1}},\
  \bibinfo {pages} {e1500608} (\bibinfo {year} {2015})}\BibitemShut {NoStop}%
\bibitem [{\citenamefont {Le~Berre}\ \emph {et~al.}(2002)\citenamefont
  {Le~Berre}, \citenamefont {Ressayre}, \citenamefont {Tallet}, \citenamefont
  {Pomeau},\ and\ \citenamefont {Di~Menza}}]{LeBerre2002}%
  \BibitemOpen
  \bibfield  {author} {\bibinfo {author} {\bibfnamefont {M.}~\bibnamefont
  {Le~Berre}}, \bibinfo {author} {\bibfnamefont {E.}~\bibnamefont {Ressayre}},
  \bibinfo {author} {\bibfnamefont {A.}~\bibnamefont {Tallet}}, \bibinfo
  {author} {\bibfnamefont {Y.}~\bibnamefont {Pomeau}}, \ and\ \bibinfo {author}
  {\bibfnamefont {L.}~\bibnamefont {Di~Menza}},\ }\href {\doibase
  10.1103/PhysRevE.66.026203} {\bibfield  {journal} {\bibinfo  {journal} {Phys.
  Rev. E}\ }\textbf {\bibinfo {volume} {66}},\ \bibinfo {pages} {026203}
  (\bibinfo {year} {2002})}\BibitemShut {NoStop}%
\bibitem [{\citenamefont {Le~Goff}\ \emph {et~al.}(2017)\citenamefont
  {Le~Goff}, \citenamefont {To},\ and\ \citenamefont
  {Pierre-Louis}}]{PierreLouis2017}%
  \BibitemOpen
  \bibfield  {author} {\bibinfo {author} {\bibfnamefont {T.}~\bibnamefont
  {Le~Goff}}, \bibinfo {author} {\bibfnamefont {T.~B.~T.}\ \bibnamefont {To}},
  \ and\ \bibinfo {author} {\bibfnamefont {O.}~\bibnamefont {Pierre-Louis}},\
  }\href {\doibase 10.1140/epje/i2017-11532-6} {\bibfield  {journal} {\bibinfo
  {journal} {The European Physical Journal E}\ }\textbf {\bibinfo {volume}
  {40}},\ \bibinfo {pages} {44} (\bibinfo {year} {2017})}\BibitemShut {NoStop}%
\bibitem [{\citenamefont {Swain}\ and\ \citenamefont
  {Andelman}(2001)}]{Swain2001}%
  \BibitemOpen
  \bibfield  {author} {\bibinfo {author} {\bibfnamefont {P.~S.}\ \bibnamefont
  {Swain}}\ and\ \bibinfo {author} {\bibfnamefont {D.}~\bibnamefont
  {Andelman}},\ }\href {\doibase 10.1103/PhysRevE.63.051911} {\bibfield
  {journal} {\bibinfo  {journal} {Phys. Rev. E}\ }\textbf {\bibinfo {volume}
  {63}},\ \bibinfo {pages} {051911} (\bibinfo {year} {2001})}\BibitemShut
  {NoStop}%
\bibitem [{\citenamefont {Sengupta}\ and\ \citenamefont
  {Limozin}(2010)}]{Sengupta2010}%
  \BibitemOpen
  \bibfield  {author} {\bibinfo {author} {\bibfnamefont {K.}~\bibnamefont
  {Sengupta}}\ and\ \bibinfo {author} {\bibfnamefont {L.}~\bibnamefont
  {Limozin}},\ }\href {\doibase 10.1103/PhysRevLett.104.088101} {\bibfield
  {journal} {\bibinfo  {journal} {Phys. Rev. Lett.}\ }\textbf {\bibinfo
  {volume} {104}},\ \bibinfo {pages} {088101} (\bibinfo {year}
  {2010})}\BibitemShut {NoStop}%
\bibitem [{\citenamefont {Israelachvili}(2015)}]{Israelachvili2015}%
  \BibitemOpen
  \bibfield  {author} {\bibinfo {author} {\bibfnamefont {J.}~\bibnamefont
  {Israelachvili}},\ }\href {https://books.google.fr/books?id=MVbWBhubrgIC}
  {\emph {\bibinfo {title} {Intermolecular and Surface Forces}}},\
  Intermolecular and Surface Forces\ (\bibinfo  {publisher} {Elsevier
  Science},\ \bibinfo {year} {2015})\BibitemShut {NoStop}%
\bibitem [{\citenamefont {Canham}(1970)}]{Canham1970}%
  \BibitemOpen
  \bibfield  {author} {\bibinfo {author} {\bibfnamefont {P.}~\bibnamefont
  {Canham}},\ }\href {\doibase https://doi.org/10.1016/S0022-5193(70)80032-7}
  {\bibfield  {journal} {\bibinfo  {journal} {Journal of Theoretical Biology}\
  }\textbf {\bibinfo {volume} {26}},\ \bibinfo {pages} {61 } (\bibinfo {year}
  {1970})}\BibitemShut {NoStop}%
\bibitem [{\citenamefont {Helfrich}(1973)}]{Helfrich1973}%
  \BibitemOpen
  \bibfield  {author} {\bibinfo {author} {\bibfnamefont {W.}~\bibnamefont
  {Helfrich}},\ }\href {\doibase https://doi.org/10.1515/znc-1973-11-1209}
  {\bibfield  {journal} {\bibinfo  {journal} {Zeitschrift für Naturforschung
  C}\ }\textbf {\bibinfo {volume} {28}},\ \bibinfo {pages} {693 } (\bibinfo
  {year} {01 Dec. 1973})}\BibitemShut {NoStop}%
\bibitem [{\citenamefont {Lipowsky}(1995)}]{Lipowsky1995}%
  \BibitemOpen
  \bibfield  {author} {\bibinfo {author} {\bibfnamefont {R.}~\bibnamefont
  {Lipowsky}},\ }\href {\doibase https://doi.org/10.1016/0959-440X(95)80040-9}
  {\bibfield  {journal} {\bibinfo  {journal} {Current Opinion in Structural
  Biology}\ }\textbf {\bibinfo {volume} {5}},\ \bibinfo {pages} {531 }
  (\bibinfo {year} {1995})}\BibitemShut {NoStop}%
\bibitem [{\citenamefont {Seifert}(1995)}]{Seifert1995}%
  \BibitemOpen
  \bibfield  {author} {\bibinfo {author} {\bibfnamefont {U.}~\bibnamefont
  {Seifert}},\ }\href {\doibase https://doi.org/10.1007/BF01307480} {\bibfield
  {journal} {\bibinfo  {journal} {Zeitschrift für Physik B Condensed Matter}\
  }\textbf {\bibinfo {volume} {97}},\ \bibinfo {pages} {299 } (\bibinfo {year}
  {1995})}\BibitemShut {NoStop}%
\bibitem [{\citenamefont {Young}\ \emph {et~al.}(2014)\citenamefont {Young},
  \citenamefont {Veerapaneni},\ and\ \citenamefont {Miksis}}]{Young2014}%
  \BibitemOpen
  \bibfield  {author} {\bibinfo {author} {\bibfnamefont {Y.-N.}\ \bibnamefont
  {Young}}, \bibinfo {author} {\bibfnamefont {S.}~\bibnamefont {Veerapaneni}},
  \ and\ \bibinfo {author} {\bibfnamefont {M.~J.}\ \bibnamefont {Miksis}},\
  }\href {\doibase 10.1017/jfm.2014.281} {\bibfield  {journal} {\bibinfo
  {journal} {Journal of Fluid Mechanics}\ }\textbf {\bibinfo {volume} {751}},\
  \bibinfo {pages} {406–431} (\bibinfo {year} {2014})}\BibitemShut {NoStop}%
\bibitem [{\citenamefont {Sheetz}(2001)}]{Sheetz2001}%
  \BibitemOpen
  \bibfield  {author} {\bibinfo {author} {\bibfnamefont {M.~P.}\ \bibnamefont
  {Sheetz}},\ }\href {\doibase https://doi.org/10.1038/35073095} {\bibfield
  {journal} {\bibinfo  {journal} {Nature Reviews Molecular Cell Biology}\
  }\textbf {\bibinfo {volume} {2}},\ \bibinfo {pages} {392} (\bibinfo {year}
  {2001})}\BibitemShut {NoStop}%
\bibitem [{\citenamefont {Speck}\ and\ \citenamefont {Vink}(2012)}]{Speck2012}%
  \BibitemOpen
  \bibfield  {author} {\bibinfo {author} {\bibfnamefont {T.}~\bibnamefont
  {Speck}}\ and\ \bibinfo {author} {\bibfnamefont {R.~L.~C.}\ \bibnamefont
  {Vink}},\ }\href {\doibase 10.1103/PhysRevE.86.031923} {\bibfield  {journal}
  {\bibinfo  {journal} {Phys. Rev. E}\ }\textbf {\bibinfo {volume} {86}},\
  \bibinfo {pages} {031923} (\bibinfo {year} {2012})}\BibitemShut {NoStop}%
\bibitem [{\citenamefont {Maciver}(1992)}]{Maciver1992}%
  \BibitemOpen
  \bibfield  {author} {\bibinfo {author} {\bibfnamefont {S.}~\bibnamefont
  {Maciver}},\ }\href {\doibase https://doi.org/10.1016/0962-8924(92)90202-X}
  {\bibfield  {journal} {\bibinfo  {journal} {Trends in Cell Biology}\ }\textbf
  {\bibinfo {volume} {2}},\ \bibinfo {pages} {282} (\bibinfo {year}
  {1992})}\BibitemShut {NoStop}%
\bibitem [{\citenamefont {Berrier}\ and\ \citenamefont
  {Yamada}(2007)}]{Berrier2007}%
  \BibitemOpen
  \bibfield  {author} {\bibinfo {author} {\bibfnamefont {A.~L.}\ \bibnamefont
  {Berrier}}\ and\ \bibinfo {author} {\bibfnamefont {K.~M.}\ \bibnamefont
  {Yamada}},\ }\href {\doibase 10.1002/jcp.21237} {\bibfield  {journal}
  {\bibinfo  {journal} {Journal of Cellular Physiology}\ }\textbf {\bibinfo
  {volume} {213}},\ \bibinfo {pages} {565} (\bibinfo {year} {2007})},\ \Eprint
  {http://arxiv.org/abs/https://onlinelibrary.wiley.com/doi/pdf/10.1002/jcp.21237}
  {https://onlinelibrary.wiley.com/doi/pdf/10.1002/jcp.21237} \BibitemShut
  {NoStop}%
\bibitem [{\citenamefont {Braga}(2002)}]{Braga2002}%
  \BibitemOpen
  \bibfield  {author} {\bibinfo {author} {\bibfnamefont {V.~M.}\ \bibnamefont
  {Braga}},\ }\href {\doibase https://doi.org/10.1016/S0955-0674(02)00373-3}
  {\bibfield  {journal} {\bibinfo  {journal} {Current Opinion in Cell Biology}\
  }\textbf {\bibinfo {volume} {14}},\ \bibinfo {pages} {546 } (\bibinfo {year}
  {2002})}\BibitemShut {NoStop}%
\bibitem [{\citenamefont {Asfaw}\ \emph {et~al.}(2006)\citenamefont {Asfaw},
  \citenamefont {R{\'{o}}{\.{z}}ycki}, \citenamefont {Lipowsky},\ and\
  \citenamefont {Weikl}}]{Asfaw2006}%
  \BibitemOpen
  \bibfield  {author} {\bibinfo {author} {\bibfnamefont {M.}~\bibnamefont
  {Asfaw}}, \bibinfo {author} {\bibfnamefont {B.}~\bibnamefont
  {R{\'{o}}{\.{z}}ycki}}, \bibinfo {author} {\bibfnamefont {R.}~\bibnamefont
  {Lipowsky}}, \ and\ \bibinfo {author} {\bibfnamefont {T.~R.}\ \bibnamefont
  {Weikl}},\ }\href {\doibase 10.1209/epl/i2006-10317-0} {\bibfield  {journal}
  {\bibinfo  {journal} {Europhysics Letters ({EPL})}\ }\textbf {\bibinfo
  {volume} {76}},\ \bibinfo {pages} {703} (\bibinfo {year} {2006})}\BibitemShut
  {NoStop}%
\bibitem [{Note1()}]{Note1}%
  \BibitemOpen
  \bibinfo {note} {The wavelength is evaluated from the simulation with the
  smallest value of the shear velocity $V$ that gives periodic
  state.}\BibitemShut {Stop}%
\bibitem [{\citenamefont {Henriksen}\ \emph {et~al.}(2006)\citenamefont
  {Henriksen}, \citenamefont {Rowat}, \citenamefont {Brief}, \citenamefont
  {Hsueh}, \citenamefont {Thewalt}, \citenamefont {Zuckermann},\ and\
  \citenamefont {Ipsen}}]{Henriksen2006}%
  \BibitemOpen
  \bibfield  {author} {\bibinfo {author} {\bibfnamefont {J.}~\bibnamefont
  {Henriksen}}, \bibinfo {author} {\bibfnamefont {A.}~\bibnamefont {Rowat}},
  \bibinfo {author} {\bibfnamefont {E.}~\bibnamefont {Brief}}, \bibinfo
  {author} {\bibfnamefont {Y.}~\bibnamefont {Hsueh}}, \bibinfo {author}
  {\bibfnamefont {J.}~\bibnamefont {Thewalt}}, \bibinfo {author} {\bibfnamefont
  {M.}~\bibnamefont {Zuckermann}}, \ and\ \bibinfo {author} {\bibfnamefont
  {J.}~\bibnamefont {Ipsen}},\ }\href {\doibase
  https://doi.org/10.1529/biophysj.105.067652} {\bibfield  {journal} {\bibinfo
  {journal} {Biophysical Journal}\ }\textbf {\bibinfo {volume} {90}},\ \bibinfo
  {pages} {1639 } (\bibinfo {year} {2006})}\BibitemShut {NoStop}%
\bibitem [{\citenamefont {Pakkanen}\ \emph {et~al.}(2011)\citenamefont
  {Pakkanen}, \citenamefont {Duelund}, \citenamefont {Qvortrup}, \citenamefont
  {Pedersen},\ and\ \citenamefont {Ipsen}}]{Pakkanen2011}%
  \BibitemOpen
  \bibfield  {author} {\bibinfo {author} {\bibfnamefont {K.~I.}\ \bibnamefont
  {Pakkanen}}, \bibinfo {author} {\bibfnamefont {L.}~\bibnamefont {Duelund}},
  \bibinfo {author} {\bibfnamefont {K.}~\bibnamefont {Qvortrup}}, \bibinfo
  {author} {\bibfnamefont {J.~S.}\ \bibnamefont {Pedersen}}, \ and\ \bibinfo
  {author} {\bibfnamefont {J.~H.}\ \bibnamefont {Ipsen}},\ }\href {\doibase
  https://doi.org/10.1016/j.bbamem.2011.04.006} {\bibfield  {journal} {\bibinfo
   {journal} {Biochimica et Biophysica Acta (BBA) - Biomembranes}\ }\textbf
  {\bibinfo {volume} {1808}},\ \bibinfo {pages} {1947 } (\bibinfo {year}
  {2011})}\BibitemShut {NoStop}%
\bibitem [{\citenamefont {Bruinsma}\ \emph {et~al.}(2000)\citenamefont
  {Bruinsma}, \citenamefont {Behrisch},\ and\ \citenamefont
  {Sackmann}}]{Bruinsma2000}%
  \BibitemOpen
  \bibfield  {author} {\bibinfo {author} {\bibfnamefont {R.}~\bibnamefont
  {Bruinsma}}, \bibinfo {author} {\bibfnamefont {A.}~\bibnamefont {Behrisch}},
  \ and\ \bibinfo {author} {\bibfnamefont {E.}~\bibnamefont {Sackmann}},\
  }\href {\doibase 10.1103/PhysRevE.61.4253} {\bibfield  {journal} {\bibinfo
  {journal} {Phys. Rev. E}\ }\textbf {\bibinfo {volume} {61}},\ \bibinfo
  {pages} {4253} (\bibinfo {year} {2000})}\BibitemShut {NoStop}%
\bibitem [{\citenamefont {Weikl}\ \emph {et~al.}(2009)\citenamefont {Weikl},
  \citenamefont {Asfaw}, \citenamefont {Krobath}, \citenamefont {Różycki},\
  and\ \citenamefont {Lipowsky}}]{Weikl2009}%
  \BibitemOpen
  \bibfield  {author} {\bibinfo {author} {\bibfnamefont {T.~R.}\ \bibnamefont
  {Weikl}}, \bibinfo {author} {\bibfnamefont {M.}~\bibnamefont {Asfaw}},
  \bibinfo {author} {\bibfnamefont {H.}~\bibnamefont {Krobath}}, \bibinfo
  {author} {\bibfnamefont {B.}~\bibnamefont {Różycki}}, \ and\ \bibinfo
  {author} {\bibfnamefont {R.}~\bibnamefont {Lipowsky}},\ }\href {\doibase
  10.1039/B902017A} {\bibfield  {journal} {\bibinfo  {journal} {Soft Matter}\
  }\textbf {\bibinfo {volume} {5}},\ \bibinfo {pages} {3213} (\bibinfo {year}
  {2009})}\BibitemShut {NoStop}%
\bibitem [{\citenamefont {Miron-Mendoza}\ \emph {et~al.}(2010)\citenamefont
  {Miron-Mendoza}, \citenamefont {Seemann},\ and\ \citenamefont
  {Grinnell}}]{Mironmendoza2010}%
  \BibitemOpen
  \bibfield  {author} {\bibinfo {author} {\bibfnamefont {M.}~\bibnamefont
  {Miron-Mendoza}}, \bibinfo {author} {\bibfnamefont {J.}~\bibnamefont
  {Seemann}}, \ and\ \bibinfo {author} {\bibfnamefont {F.}~\bibnamefont
  {Grinnell}},\ }\href {\doibase
  https://doi.org/10.1016/j.biomaterials.2010.04.064} {\bibfield  {journal}
  {\bibinfo  {journal} {Biomaterials}\ }\textbf {\bibinfo {volume} {31}},\
  \bibinfo {pages} {6425 } (\bibinfo {year} {2010})}\BibitemShut {NoStop}%
\bibitem [{\citenamefont {Schneck}\ \emph {et~al.}(2012)\citenamefont
  {Schneck}, \citenamefont {Sedlmeier},\ and\ \citenamefont
  {Netz}}]{Schneck2012}%
  \BibitemOpen
  \bibfield  {author} {\bibinfo {author} {\bibfnamefont {E.}~\bibnamefont
  {Schneck}}, \bibinfo {author} {\bibfnamefont {F.}~\bibnamefont {Sedlmeier}},
  \ and\ \bibinfo {author} {\bibfnamefont {R.~R.}\ \bibnamefont {Netz}},\
  }\href {\doibase 10.1073/pnas.1205811109} {\bibfield  {journal} {\bibinfo
  {journal} {Proceedings of the National Academy of Sciences}\ }\textbf
  {\bibinfo {volume} {109}},\ \bibinfo {pages} {14405} (\bibinfo {year}
  {2012})},\ \Eprint
  {http://arxiv.org/abs/https://www.pnas.org/content/109/36/14405.full.pdf}
  {https://www.pnas.org/content/109/36/14405.full.pdf} \BibitemShut {NoStop}%
\bibitem [{\citenamefont {Le~Goff}\ \emph {et~al.}(2014)\citenamefont
  {Le~Goff}, \citenamefont {Politi},\ and\ \citenamefont
  {Pierre-Louis}}]{LeGoff2014}%
  \BibitemOpen
  \bibfield  {author} {\bibinfo {author} {\bibfnamefont {T.}~\bibnamefont
  {Le~Goff}}, \bibinfo {author} {\bibfnamefont {P.}~\bibnamefont {Politi}}, \
  and\ \bibinfo {author} {\bibfnamefont {O.}~\bibnamefont {Pierre-Louis}},\
  }\href {\doibase 10.1103/PhysRevE.90.032114} {\bibfield  {journal} {\bibinfo
  {journal} {Phys. Rev. E}\ }\textbf {\bibinfo {volume} {90}},\ \bibinfo
  {pages} {032114} (\bibinfo {year} {2014})}\BibitemShut {NoStop}%
\end{thebibliography}%
\bibliographystyle{apsrev4-1}

\end{document}